\RequirePackage{fix-cm}
\documentclass[twocolumn,epjc3]{svjour3}  
\smartqed  
\RequirePackage{graphicx}
\journalname{Eur. Phys. J. C}
\usepackage{amssymb}
\newcommand{\be}{\begin{eqnarray}}
\newcommand{\ee}{\end{eqnarray}}
\newcommand{\eq}{\begin{eqnarray}}
\newcommand{\en}{\end{eqnarray}}

\newcommand{\bfk}{{\bf k}_{\perp}}

\newcommand{\bfb}{{\bf b}_{\perp}} 
 
\newcommand{\bfP}{{\bf P}_{\perp}} 
 
\newcommand{\bfp}{{\bf p}_{\perp}}

\begin{document}
\title{Helicity-dependent generalized parton distributions for nonzero skewness}
\author{ Chandan Mondal 
}                     
%
%
\institute{Institute of Modern Physics, Chinese Academy of Sciences, Lanzhou 730000, China.}
\date{Received: date / Revised version: date}
%
\maketitle



\begin{abstract}
We investigate the helicity dependent generalized parton distributions (GPDs) in momentum as well as transverse position (impact) spaces for $u$ and $d$ quarks in a proton when the momentum transfer in both the transverse and longitudinal directions are nonzero. The GPDs are evaluated using the light-front wavefunctions of a quark-diquark model for nucleon where the wavefunctions are constructed by the soft-wall AdS/QCD correspondence.  
We also express the GPDs in the boost-invariant longitudinal position space. 
\end{abstract}


\vskip0.2in
\noindent
\section{Introduction}
Generalized parton distributions (GPDs) play a crucial role in our understanding of the structure of the hadron in terms of the fundamental building blocks of QCD, the quarks and gluons. The GPDs (see \cite{rev,Goeke:2001tz} for reviews on GPDs) encode a wealth of information about the three dimensional spatial structure of the hadron as well as the spin and orbital angular momentum of the constituents. The GPDs are experimentally accessible in the exclusive processes like Deeply Virtual Compton Scattering (DVCS) or vector meson productions. 
At the parton level one can distinguish three kinds of parton distributions functions (PFDs): the unpolarized, the helicity distribution, and the transversity which are the functions of  longitudinal momentum faction carried by the parton($x$) only.     
The GPDs being functions of three variables $x$, square of the total momentum transferred $t$, and  the longitudinal momentum transferred $\zeta$ so called skewness in the process contain more information than the ordinary PDFs. In the forward limit, GPDs reduce to PDFs whereas the first moments of GPDs give the form factors which are accessible in exclusive processes. In parallel to three PDFs, one can define three generalized distributions namely, the unpolarized, helicity, and transversity distributions. The unpolarized and helicity GPDs are chiral-even and the transversity GPDs are chiral-odd. 
At leading twist, four chiral-even GPDs occur. Two of them are usually called unpolarized GPDs ($H$ and $E$). The other two are usually called helicity-dependent or polarized GPDs which are labeled $\widetilde{H}$ and $\widetilde{E}$. The first of them gives in forward linit the polarized quark density, the second is a spin-flip distribution which implies a change of the spin of the target.
 At zero skewness ($\zeta=0$), via Fourier transform with respect to the momentum transfer in the transverse direction ${\bf \Delta}_{\perp}$, GPDs transform to the impact parameter dependent parton distributions. Unlike the GPDs themselves, impact parameter dependent parton distributions have the interpretation of a density of partons with longitudinal momentum fraction $x$ and transverse distance $b=|\bfb|$ from the proton’s center, where $\bfb$ is the conjugate variable to ${\bf \Delta}_{\perp}$ and satisfy the positivity condition~\cite{Burk3,burk1,burk2}. The second 
moment of the GPDs corresponds to the gravitational form factors which are again related to the partonic contribution to the angular momentum  of nucleon at the $t\to 0$ limit~\cite{ji}. When one considers transversely polarized nucleons, the impact parameter dependent PDFs get distorted and the transverse distortion can also be connected with Ji's angular momentum relation. For transversely polarized state, an interesting interpretation of Ji's angular momentum sum rule~\cite{ji} was obtained in terms of the impact parameter dependent PDFs in~\cite{Burk3}. Transverse distortion arises due to the GPD $E$ for the unpolarized quark, which is related to the anomalous magnetic moment of the quarks. But in case of transversely polarized quark, the linear combination of chiral-odd GPDs ($2\widetilde{H}_T+E_T$) plays a  role similar to the GPD $E$ as for the unpolarized quark distributions.
The helicity dependent GPDs $\widetilde{H}$ in impact parameter space reflects the difference in the density of quarks with helicity equal or opposite to the proton helicity \cite{Boffi:2007yc,Diehl05,Pasquini2}.
For nonzero skewness, the GPDs can also be represented in the longitudinal position space by taking Fourier transform of the GPDs with respect to $\zeta$~\cite{CMM1,Dahiya07,BDHAV,CMM2,Kumar1,Mondal:2015uha,Chakrabarti:2015ama}.

Unlike the PFDs and form factors, it is always very difficult to measure the GPDs which can be accessed in DVCS scattering \cite{Ji:1996nm,Radyushkin:1997ki}. First experimental DVCS results in terms of the beam spin asymmetry have been presented by HERMES at DESY \cite{Airapetian:2001yk} and CLAS at JLab \cite{Stepanyan:2001sm}. Since then, much
more results are available from the measurements performed by the Hall A and Hall B/CLAS collaborations at JLab \cite{Camacho:2006qlk,Chen:2006na,Girod:2007aa,Mazouz:2007aa} and the H1, ZEUS and HERMES collaborations at DESY \cite{Chekanov:2003ya,Aktas:2005ty,Airapetian:2006zr,Aaron:2007ab,Airapetian:2011uq,Airapetian:2008aa,Airapetian:2010aa}. Exclusive production of $\omega$ meson \cite{Adolph:2016ehf}, and $\rho^0$ mesons \cite{Adolph} by scattering muons off transversely polarized proton has been measured in a very recent COMPASS experiments. The target spin asymmetries measured in these experiments agree well with GPD-based model calculations. There has been  proposals to get access to the GPDs through diffractive double meson production~\cite{Enberg,Yu}. 
The role of the GPDs in hard exclusive electroproduction of pseudoscalar mesons~\cite{Goloskokov2} as well as in leptoproduction of vector mesons~\cite{Goloskokov1} have been investigated within the framework of the handbag approach. 

Since the nonperturbative properties of hadrons are always very difficult to  evaluate from QCD first principle, there have been numerous  attempts to gain insight into the  hadron structure by studying QCD inspired models. Several theoretical predictions for the GPDs have been produced by using different descriptions of
hadron structure such as bag models \cite{Ji:1997gm,Anikin:2001zv}, soliton models \cite{Goeke:2001tz,Petrov:1998kf,Penttinen:1999th}, light-front \cite{Tiburzi:2001ta,Tiburzi:2001je,Mukherjee:2002xi}, constituent quark models (CQM) \cite{Scopetta:2003et,Scopetta:2002xq,Boffi:2002yy,Pasquini1}, AdS/QCD~\cite{Vega,CM1}. Recently, the GPDs for nonzero skewness in AdS/QCD framework has been investigated in \cite{Rinaldi:2017roc,Traini:2016jko}. In~\cite{Boffi:2003yj}, the helicity dependent GPDs for nonzero skewness in a CQM have been studied considering the Dokshitzer Gribov Lipatov Altarelli Parisi (DGLAP) region  whereas these GPDs in CQM with a kinematical range corresponding to both the DGLAP
and the Efremov Radyushkin Brodsky Lepage (ERBL) regions have been investigated in \cite{Scopetta:2004wt}.
The helicity dependent twist-two and twist-three GPDs in light-front Hamiltonian QCD for a massive dressed quark target has been presented in \cite{Mukherjee:2002xi}. The general properties of GPDs in QED 
models have been studied in both momentum and transverse position as well as longitudinal position spaces~\cite{CMM1,Mukherjee:2011an}; the impact parameter representation of the GPDs have been investigated in a QED model of a dressed electron~\cite{Dahiya07}. The moments of the GPDs have been calculated on lattice~\cite{gock,Hagler:2009,Hagler:2004,Bratt:2010}. 
In this work, we consider a light front quark-diquark model recently proposed by Gutsche et. al~\cite{Gut} where
the light-front wavefunctions are modeled from the two particle wave functions obtained in a soft-wall model of AdS/QCD correspondence~\cite{BT1,BT}. 
This model is consistent with Drell-Yan-West relation and has been shown to reproduce
many interesting nucleon properties. So far the quark-diquark model has been successfully applied to describe various aspect of nucleon properties
e.g., electromagnetic and gravitational form factor, GPDs, TMDs, charge densities, longitudinal momentum densities etc. \cite{Mondal:2015uha,Chakrabarti:2015ama,GFF,Mondal:2016xsm,Chakrabarti:2016lod,Mondal:2016afg,Chakrabarti:2016yuw,Maji:2015vsa}. More importantly, since the AdS/QCD formalism is a semiclassical approach to solve nonperturbative QCD, one can expect that the wavefunctions modeled by AdS/QCD correspondence encode the nonperturbative information of the nucleon and thus the wavefunctions are suitable to study the nonperturbative properties like GPDs, TMDs. It should be mentioned here that recently TMDs of pion have been evaluated using a model inspired by AdS/QCD correspondence \cite{Bacchetta:2017vzh}.
Here, we investigate the skewed helicity dependent GPDs in both momentum as well as transverse and longitudinal position space in this light-front quark-diquark model inspired by AdS/QCD. We also present the quark transverse distributions for $u$ and $d$ quarks in a longitudinally polarized nucleon.

The paper is organized as follows. A brief introductions about the nucleon light-front wavefunctions of quark-diquark model has been given in Section \ref{model}. In Section \ref{helicity}, we present the overlap formalism of the helicity-dependent GPDs and show the results for proton GPDs of $u$ and $d$ quarks in momentum space. The GPDs in the transverse as well as the longitudinal impact parameter space are shown in Sections \ref{helicity_trans_impact} and \ref{helicity_longi_impact}. The quark transverse distributions in the nucleon with longitudinal polarization $\Lambda(=+1)$ are presented in the Section \ref{quark_distribution}. Finally we provide a summary all the results in Section \ref{summary}.

\section{Light-front quark-diquark model constructed by AdS/QCD}\label{model}

Here we adopt the generic ansatz for the light-front quark-diquark model for the nucleons \cite{Gut} where the light-front wavefunctions
are modeled from the solution of soft-wall AdS/QCD. In this model, one contemplates the three valence quarks of the nucleons as an effective system
composed of a fermion (quark) and a composite state of diquark (boson) based on one loop quantum
fluctuations.
Then the 2-particle Fock-state expansion for proton spin components, $J^z = + \frac{1}{2}$ and   $J^z = - \frac{1}{2}$ in a frame where the transverse momentum of proton vanishes i,e. $P \equiv \big(P^+,\frac{M_n^2}{P^+},\textbf{0}_\perp\big)$, 
are written as
 \be
  |P;+\rangle 
& =& \sum_q \int \frac{dx~ d^2\textbf{k}_{\perp}}{2(2\pi)^3\sqrt{x(1-x)}}\nonumber \\
&\times&\bigg[ \psi^{+}_{+q}(x,\textbf{k}_{\perp})|+\frac{1}{2},0; xP^+,\textbf{k}_{\perp}\rangle \nonumber \\
 &&+ \psi^{+}_{-q}(x,\textbf{k}_{\perp})|-\frac{1}{2},0; xP^+,\textbf{k}_{\perp}\rangle\bigg],\\
  |P;-\rangle 
& =& \sum_q \int \frac{dx~d^2\textbf{k}_{\perp}}{2(2\pi)^3\sqrt{x(1-x)}}\nonumber \\
&\times&\bigg[\psi^{-}_{+q}(x,\textbf{k}_{\perp})|+\frac{1}{2},0; xP^+,\textbf{k}_{\perp}\rangle \nonumber \\
 &+& \psi^{-}_{-q}(x,\textbf{k}_{\perp})|-\frac{1}{2},0; xP^+,\textbf{k}_{\perp}\rangle\bigg].
  \ee
However, for nonzero transverse momentum of proton, i.e. $\bfP\ne0$, the physical transverse momenta of quark and diquark are $\bfp^q=x\bfP+\bfk$ and $\bfp^D=(1-x)\bfP-\bfk$, respectively, where $\bfk$ represents the relative transverse momentum of the constituents. $\psi_{\lambda_q q}^{\lambda_N}(x,\bfk)$ are the light-front wavefunctions with nucleon helicities $\lambda_N=\pm$ and for the struck quark $\lambda_q=\pm$; plus and minus
correspond to $+\frac{1}{2}$ and $-\frac{1}{2}$ respectively. The light-front wavefunctions are given by \cite{Gut} 
\be\label{WF}
\psi_{+q}^+(x,\bfk) &=&  \varphi_q^{(1)}(x,\bfk) \,,\nonumber\\ 
\quad
\psi_{-q}^+(x,\bfk) &=& -\frac{k^1 + ik^2}{xM_n}   \, \varphi_q^{(2)}(x,\bfk) \,, \nonumber\\
\psi_{+q}^-(x,\bfk) &=& \frac{k^1 - ik^2}{xM_n}  \, \varphi_q^{(2)}(x,\bfk)\,. \\
\psi_{-q}^-(x,\bfk) &=& \varphi_q^{(1)}(x,\bfk),\nonumber
\ee
Here, $\varphi_q^{(i=1,2)}(x,\bfk)$ are the modified wave functions which are constructed by soft-wall
AdS/QCD, after introducing the parameters $a_q^{(i)}$ and $b_q^{(i)}$ for quark $q$,
\be\label{wf2}
\varphi_q^{(i)}(x,\bfk)&=&N_q^{(i)}\frac{4\pi}{\kappa}\sqrt{\frac{\log(1/x)}{1-x}}x^{a_q^{(i)}}
(1-x)^{b_q^{(i)}}\nonumber \\
&\times&\exp\bigg[-\frac{\bfk^2}{2\kappa^2}\frac{\log(1/x)}{(1-x)^2}\bigg].
\ee
$\varphi_q^{(i)}(x,\bfk)$ reduces to the AdS/QCD solution when $a_q^{(i)}=b_q^{(i)}=0$ 
\cite{BT}. In this work, we take the AdS/QCD scale parameter $\kappa =0.4$ GeV, obtained by
fitting the nucleon form factors in the soft-wall model of AdS/QCD \cite{CM1,CM2}. The
parameters $a^{(i)}_q$ and $b^{(i)}_q$ with the constants $N^{(i)}_q$ are obtained by
fitting the electromagnetic properties of the nucleons: $F_1^q(0)=n_q$ and $F_2^q(0)=\kappa_q$
where $n_u=2$ and $n_d=1$, the number of valence $u$ and $d$ quarks in proton and
the anomalous magnetic moments for the $u$ and $d$ quarks are $\kappa_u=1.673$ and
$\kappa_d=-2.033$ \cite{Chakrabarti:2015ama}. The parameters are given by $a^{(1)}_u  = 0.020,~  a^{(1)}_d= 0.10,~
b^{(1)}_u = 0.022,~b^{(1)}_d=0.38,~
a^{(2)}_u=  1.05,~ a^{(2)}_d=  1.07,~
b^{(2)}_u= -0.15, ~b^{(2)}_d= -0.20,
N^{(1)}_u = 2.055,~ N^{(1)}_d = 1.7618,
N^{(2)}_u= 1.322, N^{(2)}_d = -2.4827 $.

\section{Helicity dependent generalized parton distributions}\label{helicity}
The helicity dependent GPDs are defined as off-forward matrix elements of the bilocal operator of light-front correlation functions of the 
axial vector current~\cite{rev,ji,diehl01}
\be
\label{gpd_eq}
 &&\frac{1}{2} \int \frac{d z^-}{2\pi}\, e^{ix P^+ z^-}\nonumber\\
&\times&  \langle p',\lambda'|\, 
     \bar{\psi}(-{\textstyle\frac{1}{2}}z)\, 
     \gamma^+ \gamma_5\, \psi({\textstyle\frac{1}{2}}z)\, 
  \,|p,\lambda \rangle \Big|_{z^+=0,\, \mathbf{z}_T=0}  
\nonumber \\
&=& \frac{1}{2P^+} \bar{u}(p',\lambda') \left[
  \widetilde{H}^q\, \gamma^+ \gamma_5 +
  \widetilde{E}^q\, \frac{\gamma_5 \Delta^+}{2M}
  \right] u(p,\lambda),
\ee
where $p$ $(p')$ and $\lambda$ $(\lambda')$ denote the proton momenta and the helicity of the initial (final) state of proton, respectively. 
The kinematical variables in the symmetric frame are  
\be
P^\mu=\frac{(p+p')^\mu}{2}, ~~ \Delta^\mu=p'^\mu-p^\mu, ~~ \zeta=-\Delta^+/2P^+,
\ee
and $t=\Delta^2$. For $\zeta=0$, $t=-{\bf\Delta}_\perp^2$. We work in the light-front gauge $A^+=0$, so that the gauge link appearing in between the quark fields in Eq.(\ref{gpd_eq}) is unity. The quark helicity conserving distributions can be related to the following matrix elements \cite{diehl01,Boffi:2007yc}
\be
  \label{amplitude}
A_{\lambda'+, \lambda +} &=&
\int \frac{d z^-}{2\pi}\, e^{i\bar x P^+ z^-}  \langle p',\lambda'|\, {\cal O}_{+,+}(z)
  \,|p,\lambda \rangle \Big|_{z^+=\vec{z}_\perp=0} \, ,
\nonumber \\
A_{\lambda'-, \lambda -} &=&
\int \frac{d z^-}{2\pi}\, e^{i\bar x P^+ z^-}  \langle p',\lambda'|\, {\cal O}_{-,-}(z)
  \,|p,\lambda \rangle \Big|_{z^+=\vec{z}_\perp=0} \, ,
\ee
where the operators $O_{+,+}$ and $O_{-,-}$ occurring in the
definitions of the quark distributions are
\be
{\cal O}_{+,+} 
&=& \frac{1}{4}\, 
  \bar{\psi}\, \gamma^+ (1+\gamma_5)\, \psi \,,\nonumber\\
{\cal O}_{-,-} 
&=& \frac{1}{4}\, 
  \bar{\psi}\, \gamma^+ (1-\gamma_5)\, \psi.
\hspace{2em}
\label{eq:op_o}
\ee
One can explicitly derive the following relations in the reference frame where the momenta $\vec p$ and $\vec p\,'$ lie in the $x-z$ plane~\cite{diehl01}
\be \label{relations}
A_{++,++} &=& \sqrt{1-\zeta^2} \left( \frac{H^q+\widetilde{H}^q}{2} - 
            \frac{\zeta^2}{1-\zeta^2}\, \frac{E^q+\widetilde{E}^q}{2} \right) , 
\nonumber \\
A_{-+,-+} &=& \sqrt{1-\zeta^2} \left( \frac{H^q-\widetilde{H}^q}{2} - 
            \frac{\zeta^2}{1-\zeta^2}\, \frac{E^q-\widetilde{E}^q}{2} \right) , 
\nonumber \\
A_{++,-+} &=& - \epsilon\,
              \frac{\sqrt{t_0-t}}{2m}\, \frac{E^q-\zeta\widetilde{E}^q}{2} , 
\nonumber \\
A_{-+,++} &=& \epsilon\,
              \frac{\sqrt{t_0-t}}{2m}\, \frac{E^q+\zeta\widetilde{E}^q}{2} ,
\ee
where, $\epsilon =
\mathrm{sgn}(D^1)$, and $D^1$ is the $x$-component of $D^\alpha =
P^+ \Delta^\alpha - \Delta^+ P^\alpha$ where $D^1=0$ corresponds to $t=t_0$. For given $\zeta$, the minimum value of $-t$ is $- t_0 = 4 m^2 \zeta^2/(1-\zeta^2)$. Due to parity invariance, one has the relations :
$
A_{-\lambda'-\mu', -\lambda-\mu} = (-1)^{\lambda'-\mu'-\lambda+\mu}\,
                                   A_{\lambda'\mu', \lambda\mu}
$
for definite quark helicities $\mu$ and $\mu'$. We can now compute the helicity dependent GPDs $\widetilde{H}^q$ and $\widetilde{E}^q$ using the relations in Eq.(\ref{relations}) as
\be
\widetilde{H}^q&=&\frac{1}{\sqrt{1-\zeta^2}}T^q_1+\frac{2M\zeta} {\sqrt{t_0 - t}(1 - \zeta^2)}T^q_2,\\
\widetilde{E}^q&=&\frac{2M} {\epsilon \zeta\sqrt{t_0 - t}}T^q_2,\label{helicity_gpds}
\ee
where the matrix elements $T^q_i$, in terms of the quark helicity basis are given by
\be
&&T^q_{1} =A_{++,++} - A_{-+,-+}, \nonumber \\
&&T^q_{2} =A_{++,-+} + A_{-+,++}.
\ee
\subsection{Overlap formalism}
\begin{figure*}[htbp]
\begin{minipage}[c]{0.98\textwidth}
\small{(a)}
\includegraphics[width=7.5cm,height=5.15cm,clip]{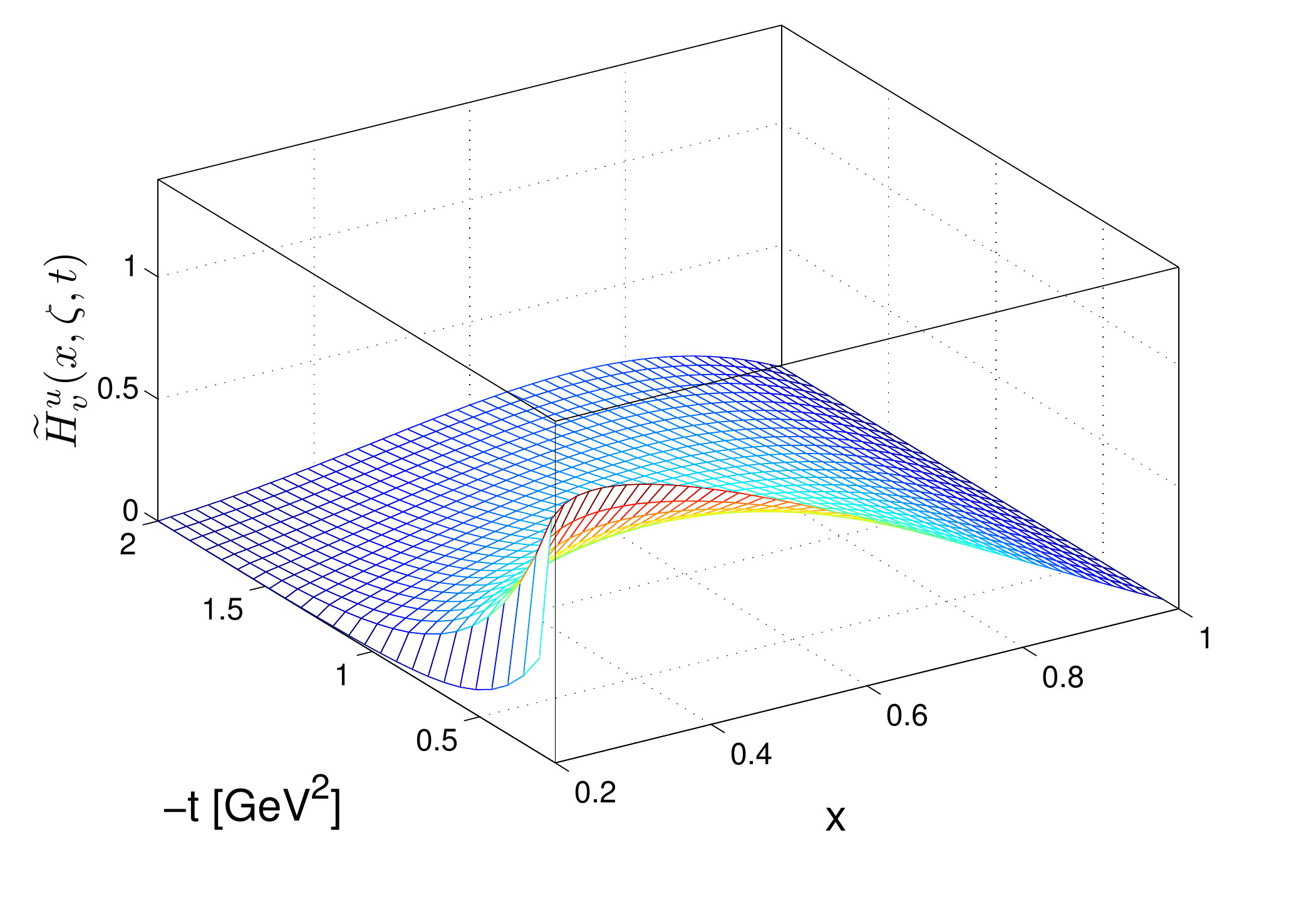}
\hspace{0.1cm}%
\small{(b)}\includegraphics[width=7.5cm,height=5.15cm,clip]{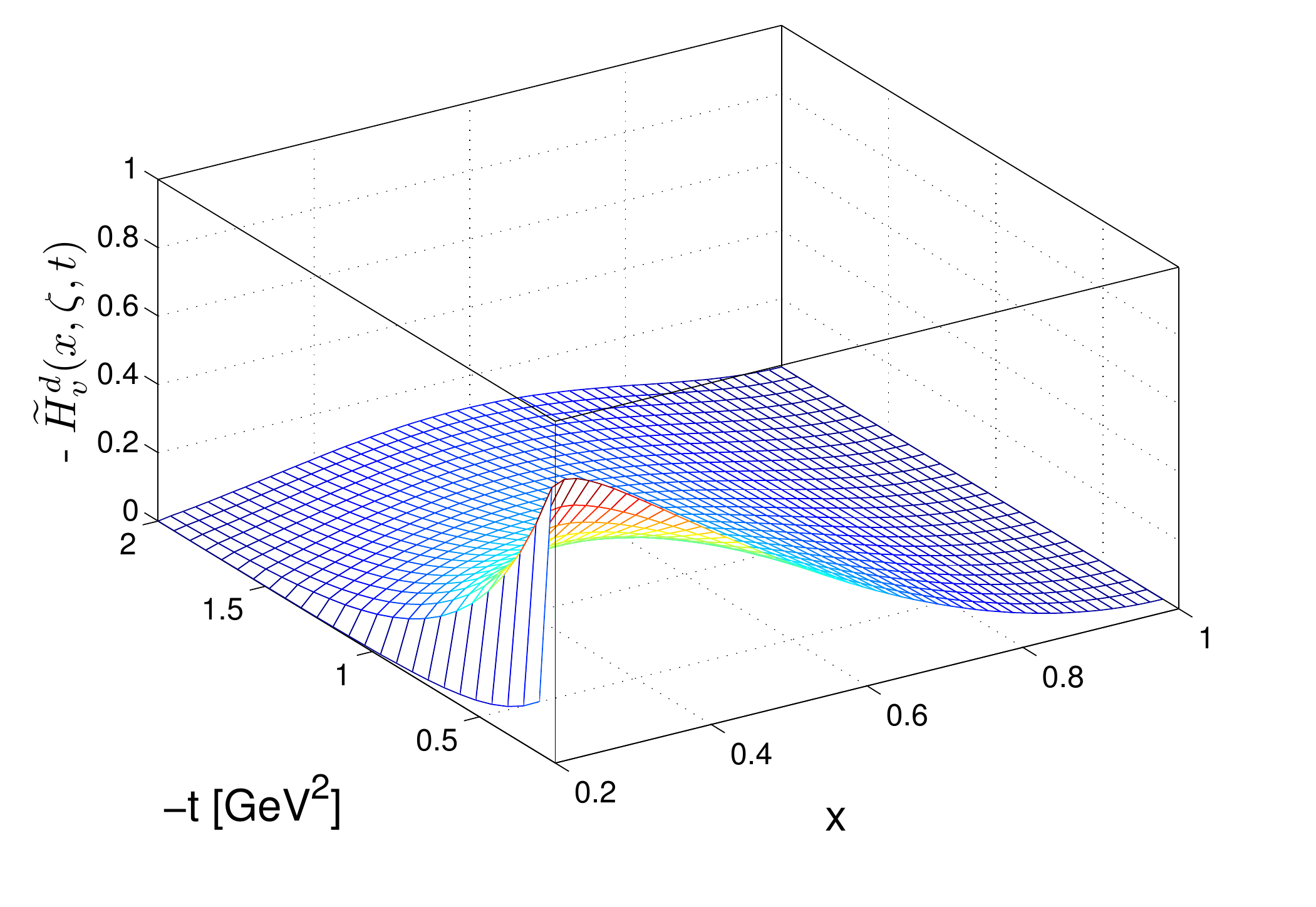}
\end{minipage}
\begin{minipage}[c]{0.98\textwidth}
\small{(c)}\includegraphics[width=7.5cm,height=5.15cm,clip]{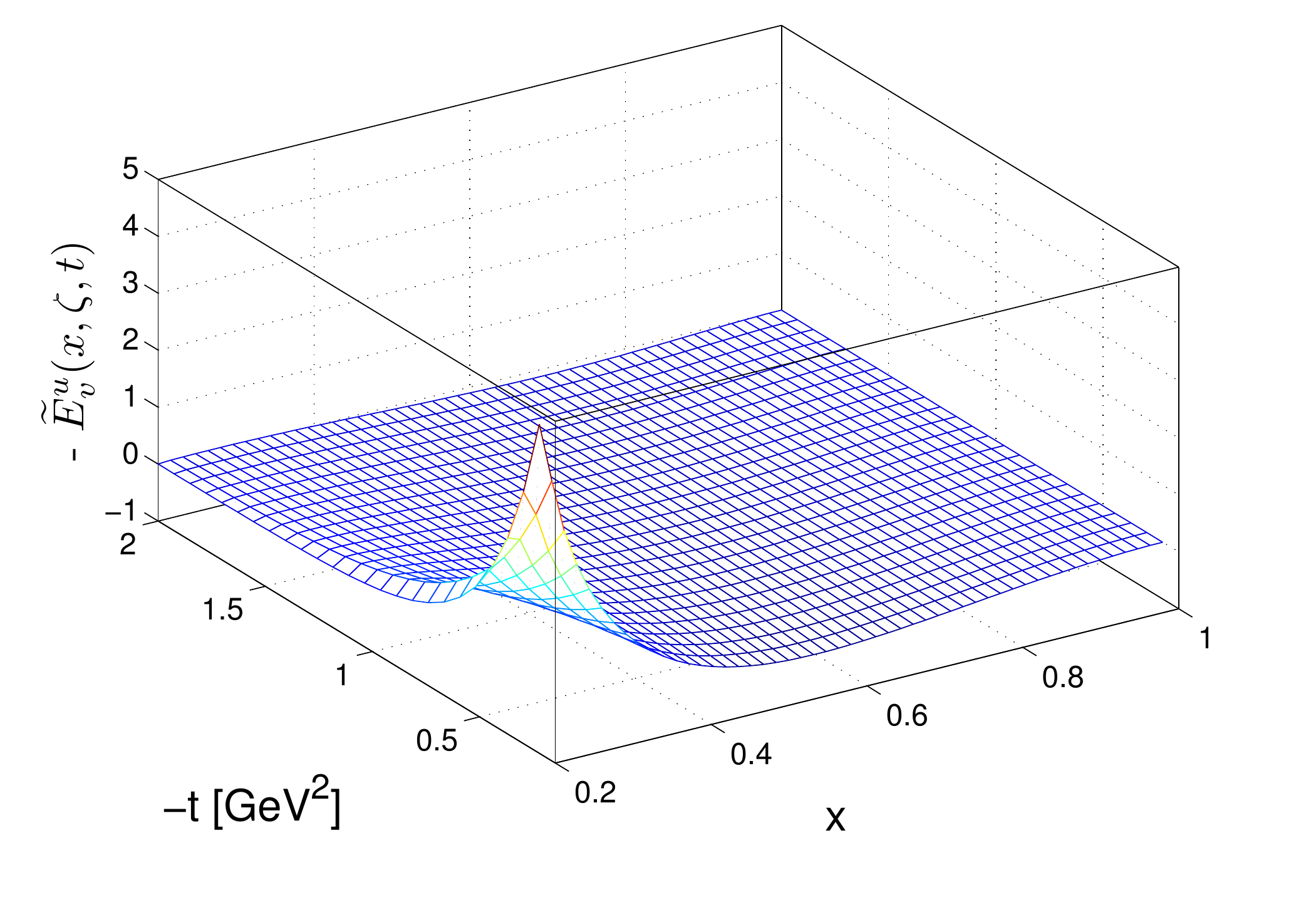}
\hspace{0.1cm}%
\small{(d)}\includegraphics[width=7.5cm,height=5.15cm,clip]{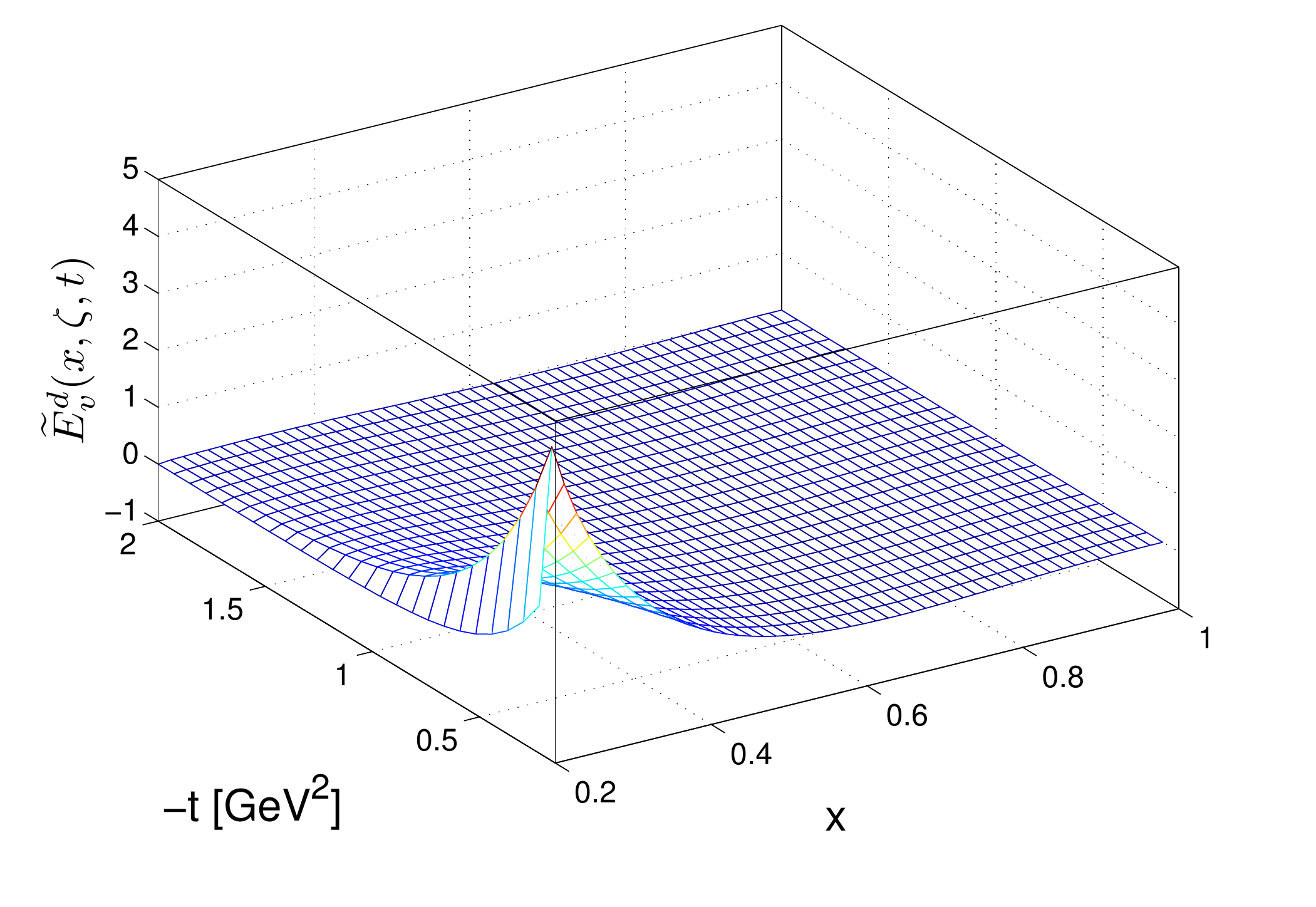}
\end{minipage}
\caption{\label{gz2}(Color online) Plots of helicity dependent GPDs for the nonzero skewness as functions of $x$ and $-t$, and for fixed value of $\zeta=0.2$. (a) $\widetilde{H}^u_v$, (b) $\widetilde{H}^d_v$ and (c) $\widetilde{E}^u_v$, (d) $\widetilde{E}^d_v$ ; for $\zeta=0.2$ the minimum value of $-t=-t_0=0.147$ $\rm GeV^2$.}
\end{figure*}
\begin{figure*}[htbp]
\begin{minipage}[c]{0.98\textwidth}
\small{(a)}
\includegraphics[width=7.5cm,height=5.15cm,clip]{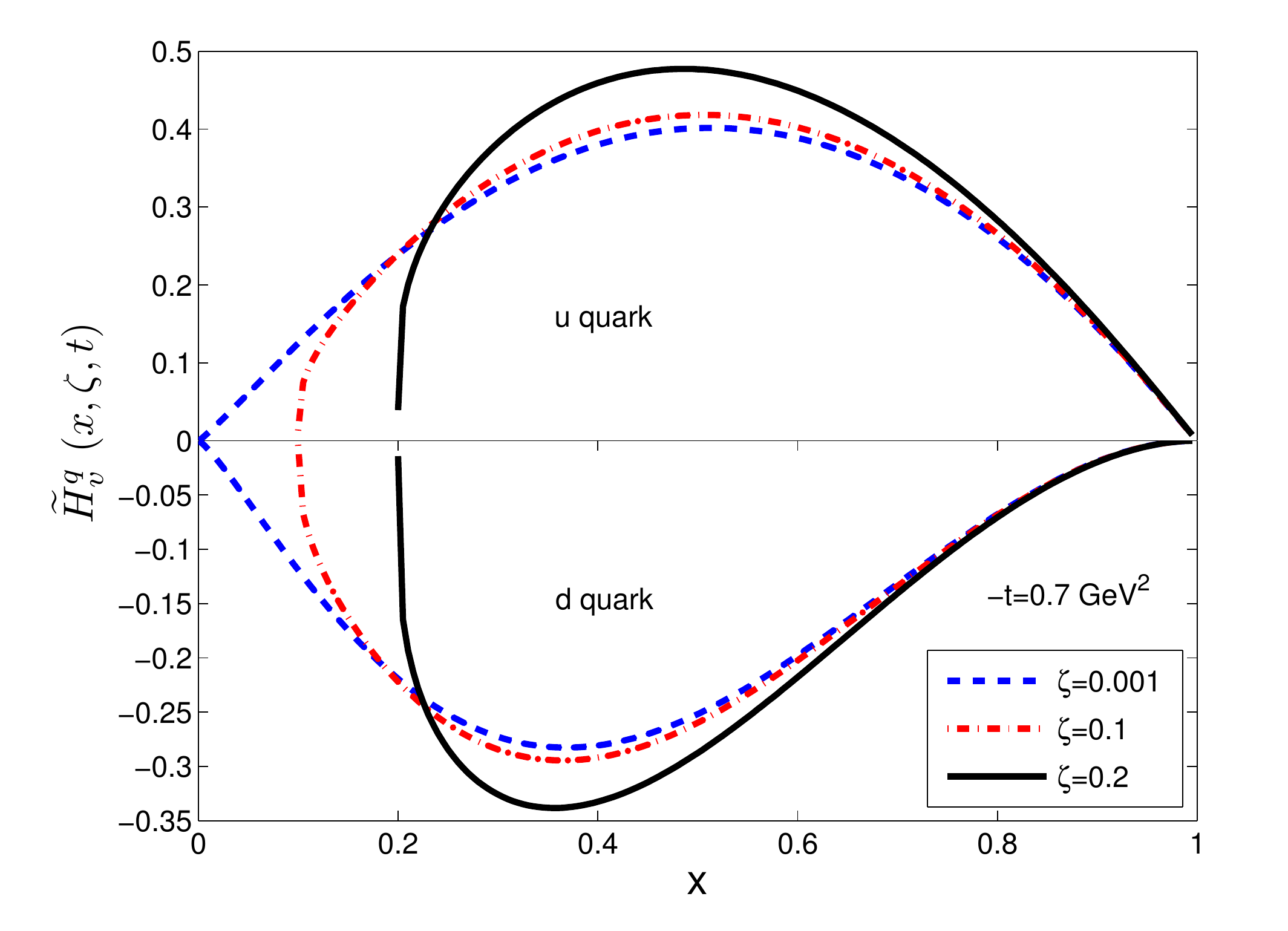}
\hspace{0.1cm}%
\small{(b)}\includegraphics[width=7.5cm,height=5.15cm,clip]{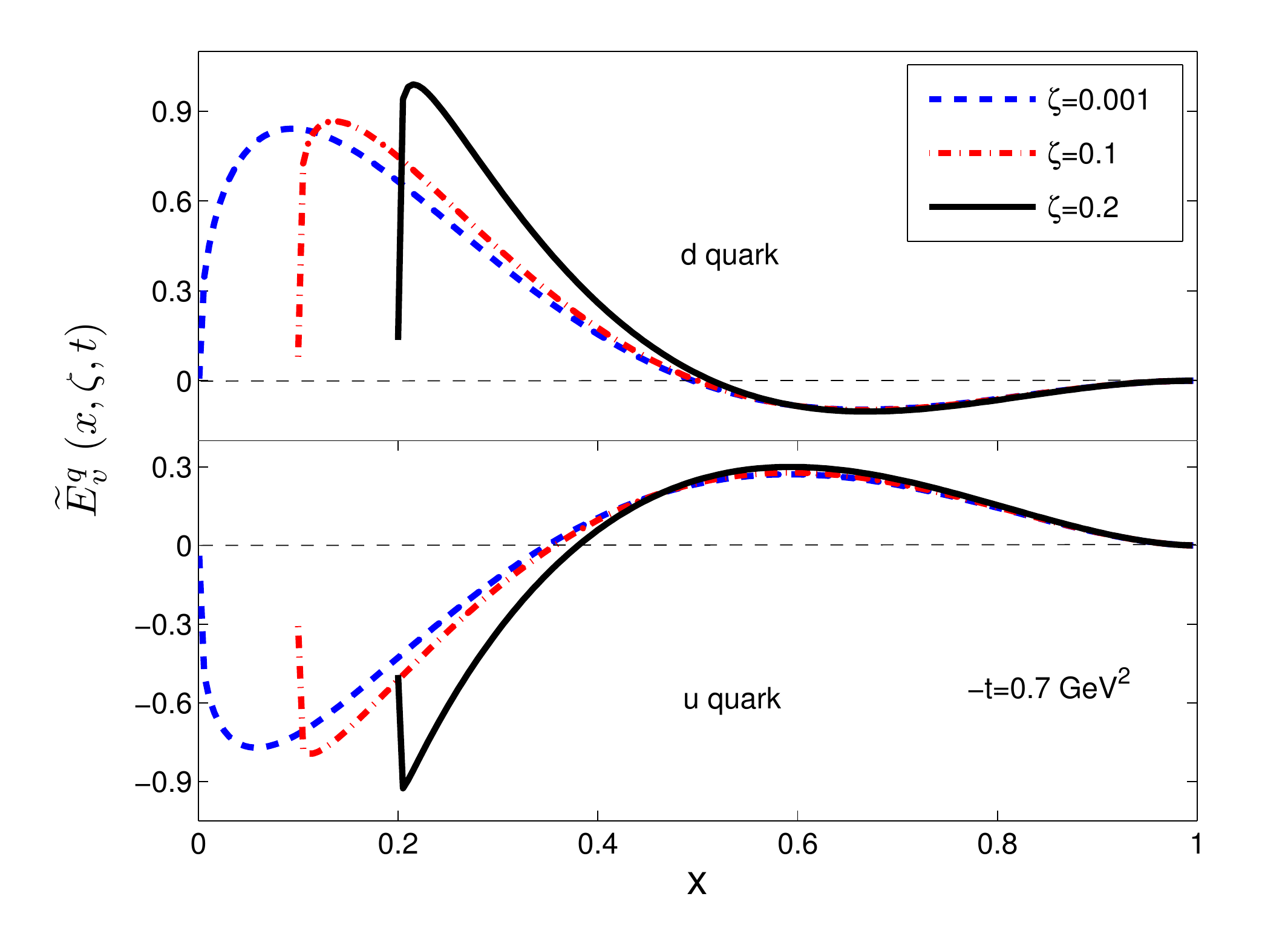}
\end{minipage}
\caption{\label{gt7}(Color online) Plots of helicity dependent GPDs for the nonzero skewness vs $x$ and different values of $\zeta$, for fixed value of $t=-0.7$ $GeV^2$. (a) $\widetilde{H}^q_v$ and (b) $\widetilde{E}^q_v$ ; $q$ stands for $u$ and $d$ quark.}
\end{figure*}
\begin{figure*}[htbp]
\begin{minipage}[c]{0.98\textwidth}
\small{(a)}
\includegraphics[width=7.5cm,height=5.15cm,clip]{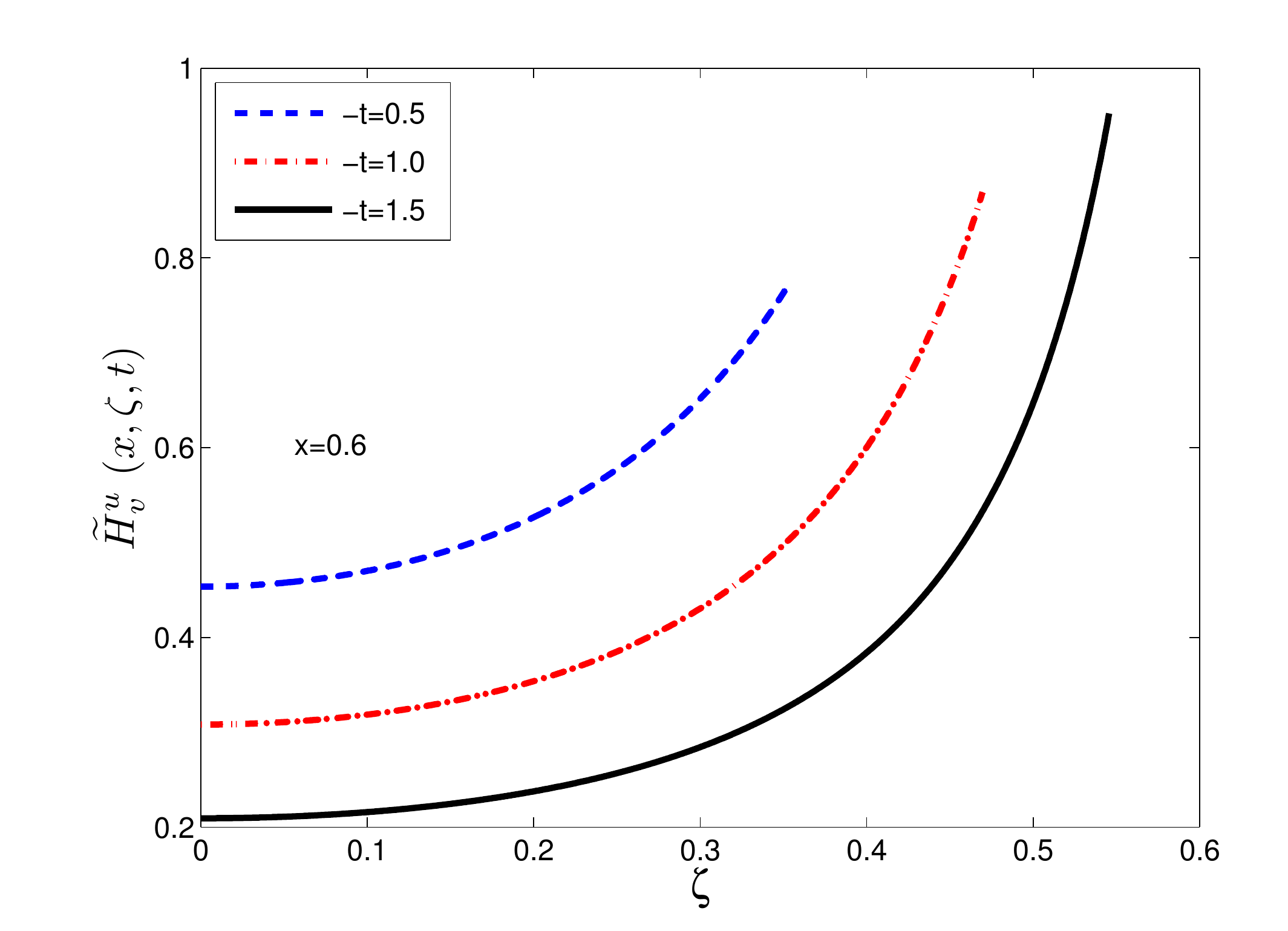}
\hspace{0.1cm}%
\small{(b)}\includegraphics[width=7.5cm,height=5.15cm,clip]{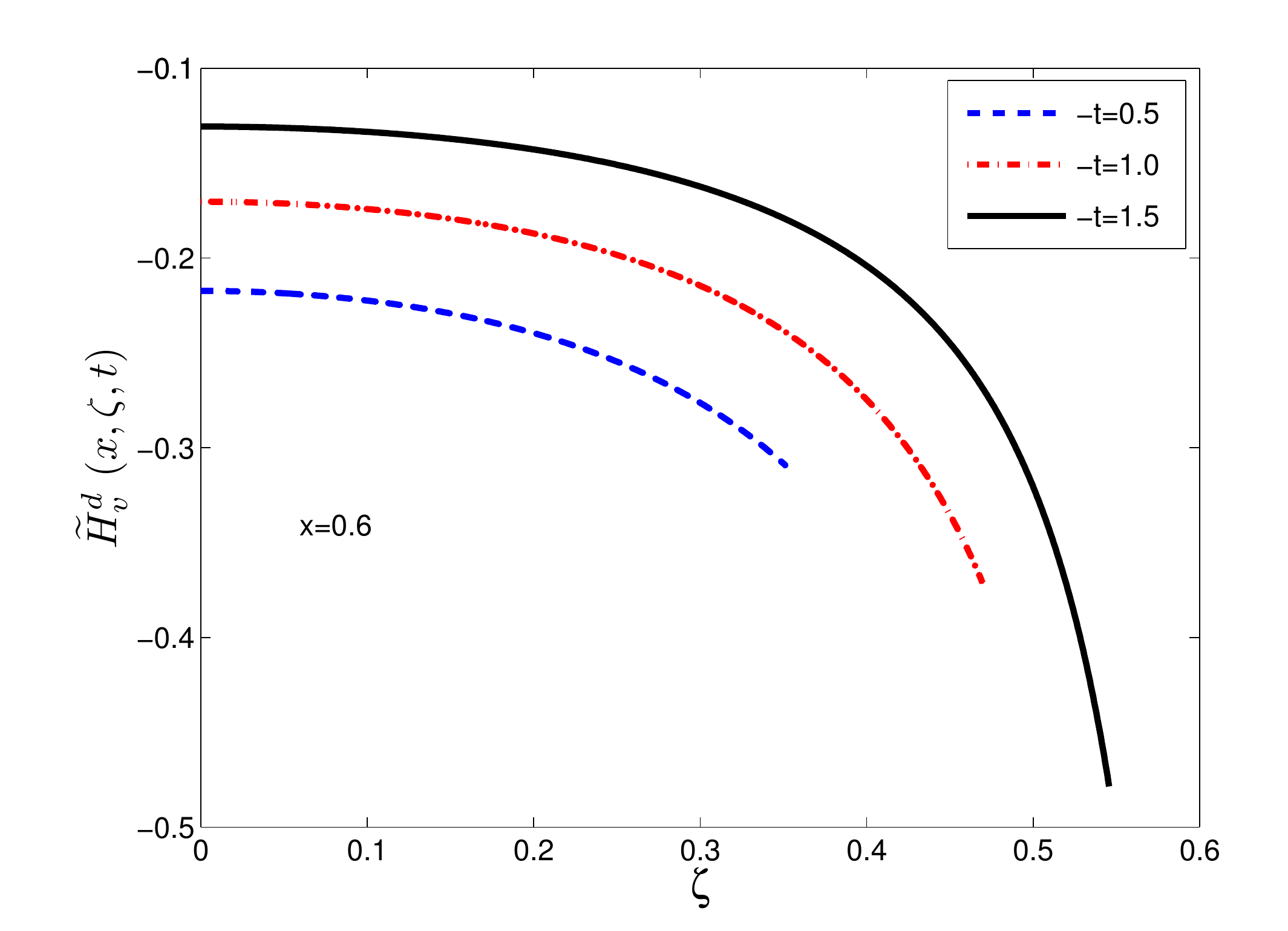}
\end{minipage}
\begin{minipage}[c]{0.98\textwidth}
\small{(c)}\includegraphics[width=7.5cm,height=5.15cm,clip]{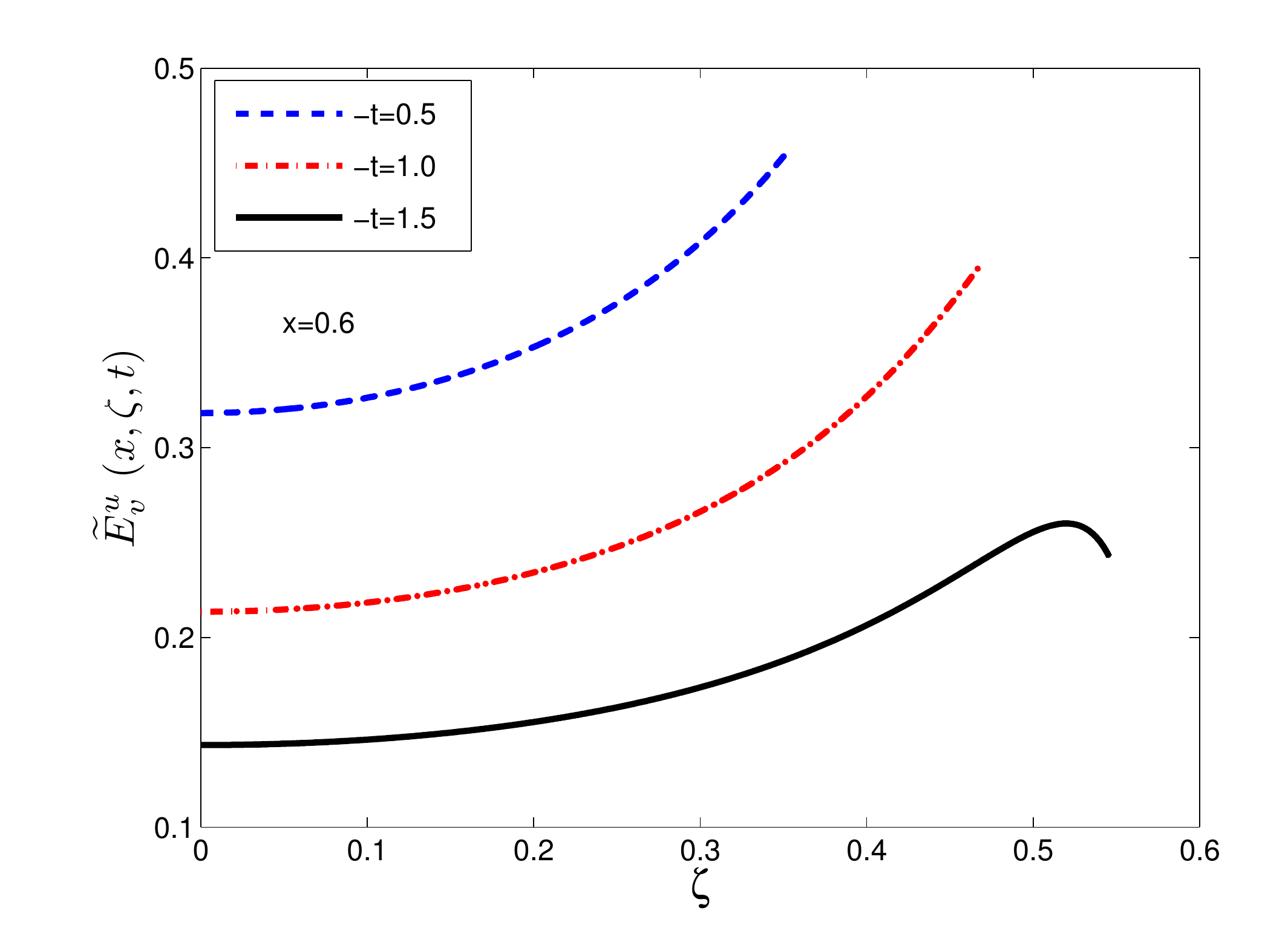}
\hspace{0.1cm}%
\small{(d)}\includegraphics[width=7.5cm,height=5.15cm,clip]{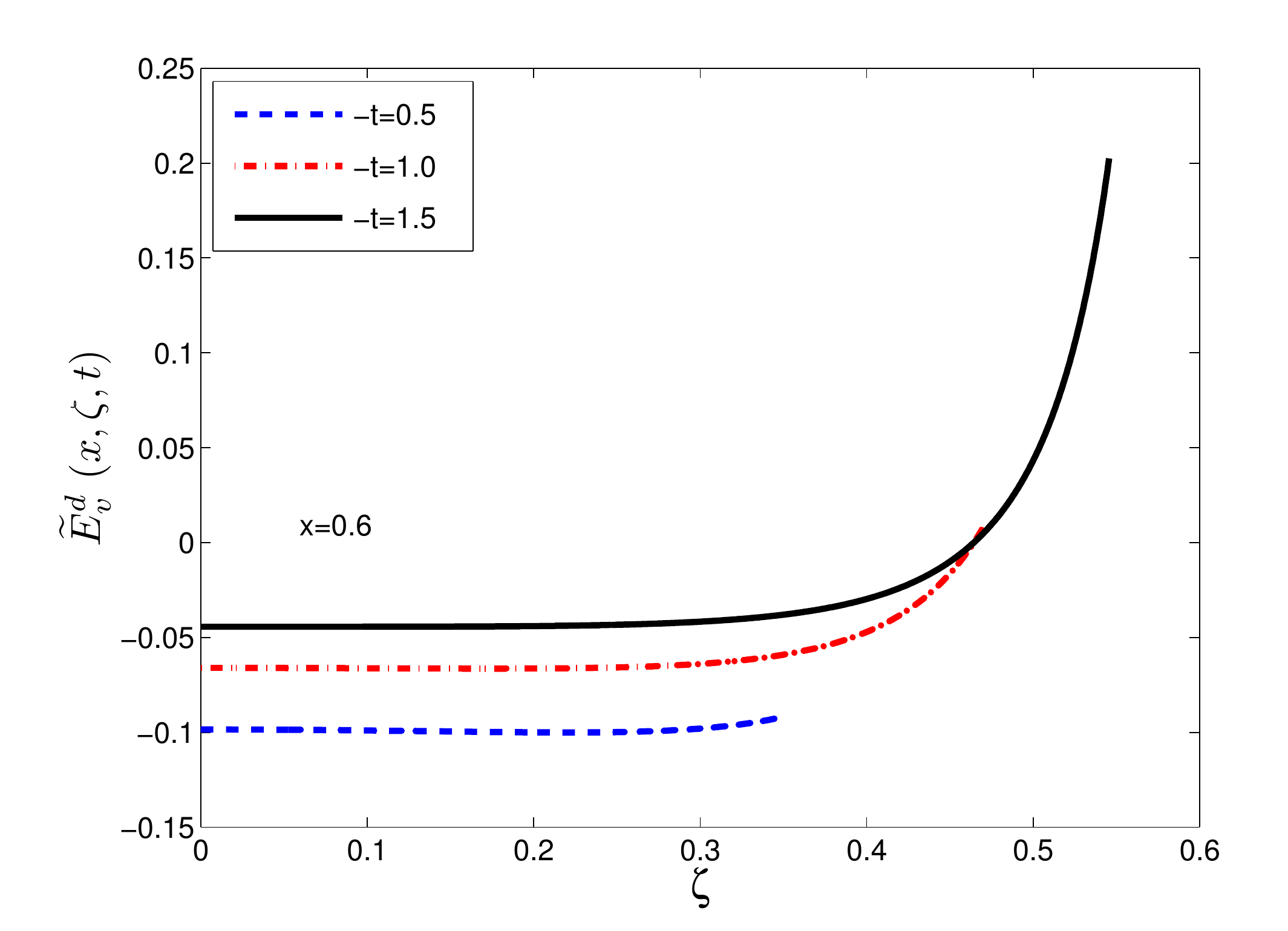}
\end{minipage}
\caption{\label{gx6}(Color online) Plots of helicity dependent GPDs for the nonzero skewness vs $\zeta$ and different values of $-t$ in $GeV^2$, for fixed value of $x=0.6$. Left panel is for $u$ quark and the right panel is for $d$ quark.}
\end{figure*}
We evaluate the helicity dependent GPDs in light front quark-diquark model using the overlap representation of light front wave functions. We consider the DGLAP region for our discussion. This kinematical domain i.e., $\zeta<x<1$ where $x$ is the light front longitudinal momentum fraction carried by the struck quark and $\zeta$ is the skewness, corresponds to the situation where one removes a quark from the initial proton  with light-front longitudinal momentum $(x+\zeta)P^+$ and re-insert it into the final proton with longitudinal momentum $(x-\zeta)P^+$. The particle number remain conserved in this kinematical region which describes the diagonal $n\rightarrow n$ overlaps. The matrix elements $T^{q}_{i}$ in the diagonal $2\rightarrow 2$ overlap representation, in terms of light-front wave functions in the quark-diquark model are given by
\be
T^q_{1}&=& \int \frac{d^2\bfk}{16\pi^3}~\bigg[\psi_{+q}^{+*}(x',\bfk')\psi_{+q}^+(x'',\bfk'') \nonumber \\
&&-\psi_{+q}^{-*}(x',\bfk')\psi_{+q}^-(x'',\bfk'')\bigg],\label{T1}\\
T^q_{2}&=& \int \frac{d^2\bfk}{16\pi^3}~\bigg[\psi_{+q}^{+*}(x',\bfk')\psi_{+q}^-(x'',\bfk'') \nonumber \\
&&+\psi_{+q}^{-*}(x',\bfk')\psi_{+q}^+(x'',\bfk'')\bigg],\label{T2}
\ee
where, for the final struck quark
\be
x'=\frac{x-\zeta}{1-\zeta}, \quad\quad\quad \bfk'=\bfk+(1-x')\frac{\bf{\Delta}_{\perp}}{2},
\ee
and for the initial struck quark
\be
x''=\frac{x+\zeta}{1+\zeta}, \quad\quad\quad \bfk''=\bfk-(1-x'')\frac{\bf{\Delta}_{\perp}}{2}.
\ee
Using the light front wave functions of the quark-diquark model given in Eq.(\ref{WF}), the explicit calculation of the matrix elements $T^{q}_{i}$ gives 
\be
T^q_{1}(x,\zeta,t)&=&\Delta q\frac{\mathcal{T}^q_1(x,\zeta,t)}{\mathcal{I}(0)},\nonumber\\
T^q_{2}(x,\zeta,t)&=&\mathcal{T}^q_2(x,\zeta,t).
\ee
with $\mathcal{I}(0)=\int_0^1 dx ~\mathcal{T}^q_1(x,0,0)$, and $\Delta q$ is the axial charge of quark $q$. The functions $\mathcal{T}^q_i(x,\zeta,t)$ are given by
\be
\mathcal{T}^q_{1}&=&\frac{1}{\kappa^2}\Big[{\frac{\log x'\log x''}{(1-x')(1-x'')}}\Big]^{1/2}\bigg[(N_q^{(1)})^2(x'x'')^{a_q^{(1)}}\times\nonumber \\
&&\{(1-x')(1-x'')\}^{b_q^{(1)}}\frac{1}{A}
-(N_q^{(2)})^2\frac{1}{M_n^2}(x'x'')^{a_q^{(2)}-1}\nonumber \\
&\times&\{(1-x')(1-x'')\}^{b_q^{(2)}}
\Big\{\frac{1}{A^2}+\Big(\frac{B^2}{4A^2}-\frac{1}{4}(1-x')\nonumber \\
&\times&(1-x'')+\frac{B}{4A}(x''-x')\Big)\frac{Q^2}{A}\Big\}\bigg]\nonumber \\
&\times&\exp\Big[Q^2\Big(C+\frac{B^2}{4A}\Big)\Big] ,\label{T11}\\
&&\nonumber\\
\mathcal{T}^q_{2}&=&\frac{N_q^{(1)}N_q^{(2)}}{\kappa^2}\Big[{\frac{\log x'\log x''}{(1-x')(1-x'')}}\Big]^{1/2}\frac{1}{M_n}\bigg[(x')^{a_q^{(1)}}\nonumber \\
&\times&(1-x')^{b_q^{(1)}}(x'')^{a_q^{(2)}-1}(1-x'')^{b_q^{(2)}}\Big(\frac{BQ}{2A^2}\nonumber \\
&-&\frac{Q}{2A}(1-x'')\Big)+(x')^{a_q^{(2)}-1}(1-x')^{b_q^{(2)}}(x'')^{a_q^{(1)}}\nonumber \\
&\times&(1-x'')^{b_q^{(1)}}
\Big(\frac{BQ}{2A^2}+\frac{Q}{2A}(1-x')\Big)\bigg]\nonumber \\
&\times&\exp\Big[Q^2\Big(C+\frac{B^2}{4A}\Big)\Big] \label{T22},
\ee
where ${{\Delta}}_{\perp}^2=Q^2=-t(1-\zeta^2)-4M_n^2\zeta^2$. $A$, $B$ and $C$ are  functions of $x'$ and $x''$,
\be
A&=&A(x,x')=-\frac{\log x'}{2\kappa^2(1-x)^2}-\frac{\log x''}{2\kappa^2(1-x')^2},\nonumber\\
B&=&B(x,x')=\frac{\log x'}{2\kappa^2(1-x)}-\frac{\log x''}{2\kappa^2(1-x')},\\
C&=&C(x,x')=\frac{1}{4}\Big[\frac{\log x'}{2\kappa^2}+\frac{\log x''}{2\kappa^2}\Big].\nonumber
\ee
Using the matrix elements calculated in Eqs.(\ref{T1}-\ref{T2}) we compute the helicity dependent GPDs in Eq.(\ref{helicity_gpds}). The GPD $\widetilde{H}^q$ are suitably normalize by the axial charge $\Delta q$ where the experimental values of $\Delta u=0.82$, and $\Delta d=-0.45$ \cite{Lorce:2014mxa,Leader:2010rb}.


The helicity dependent GPDs for nonzero skewness ($\zeta\ne 0$) for $u$ and $d$ quarks are shown in Fig.\ref{gz2}-\ref{gx6}. In Fig.\ref{gz2}, the GPDs are shown as functions of $x$ and $-t$ and a fixed value of $\zeta=0.2$ whereas in Fig.\ref{gt7}, we plot the GPDs for fixed value of $-t=0.7~ \rm{GeV^2}$ but different values of $\zeta$. One can notice that the height of the peaks of the distributions increase and move to higher $x$ with increasing $\zeta$ for fixed $-t$. The GPDs fall to zero at $x=\zeta$ when $\zeta$ is very low or the value of $-t$ is high. The reason is that in our approach we consider the contribution only from the valence quarks. Since the quark-diquark model itself depends only on the valence quarks, we can not evaluate the total (sea+valence) GPDs in this model.  
The similar behavior of the helicity dependent GPDs has been found in the relativistic constituent quark model calculated in~\cite{Boffi:2003yj}. Also, the ERBL region, i.e. $x<\zeta$ where quark-antiquark pair creation/annihilation are involved is not included in this model.
In Fig.\ref{gx6}, we show the GPDs as functions of $\zeta$ for fixed $x$ and different values of $-t$. The GPDs rise smoothly as $\zeta$ increases for all $t$ values and GPDs have  different values at $\zeta=0$ for different values of $-t$. The similar behaviors have also been observed for the unpolarized and chiral-odd GPDs (except $\widetilde{E}_T$, it is odd in $\zeta$) in quark-diquark model \cite{Mondal:2015uha,Chakrabarti:2015ama}, phenomenological QED model \cite{CMM1}. It can also be noticed that $\widetilde{E}^u(x,\zeta,t)$ shows markedly different behavior from the other GPDs. $\widetilde{E}^u(x,\zeta,t)$ rises smoothly as $\zeta$ increases but the magnitude at $\zeta_{max}=\sqrt{(-t)/(-t+4M_n^2)}$ decreases with increasing $-t$.

\begin{figure}[htbp]
\includegraphics[width=8cm,clip]{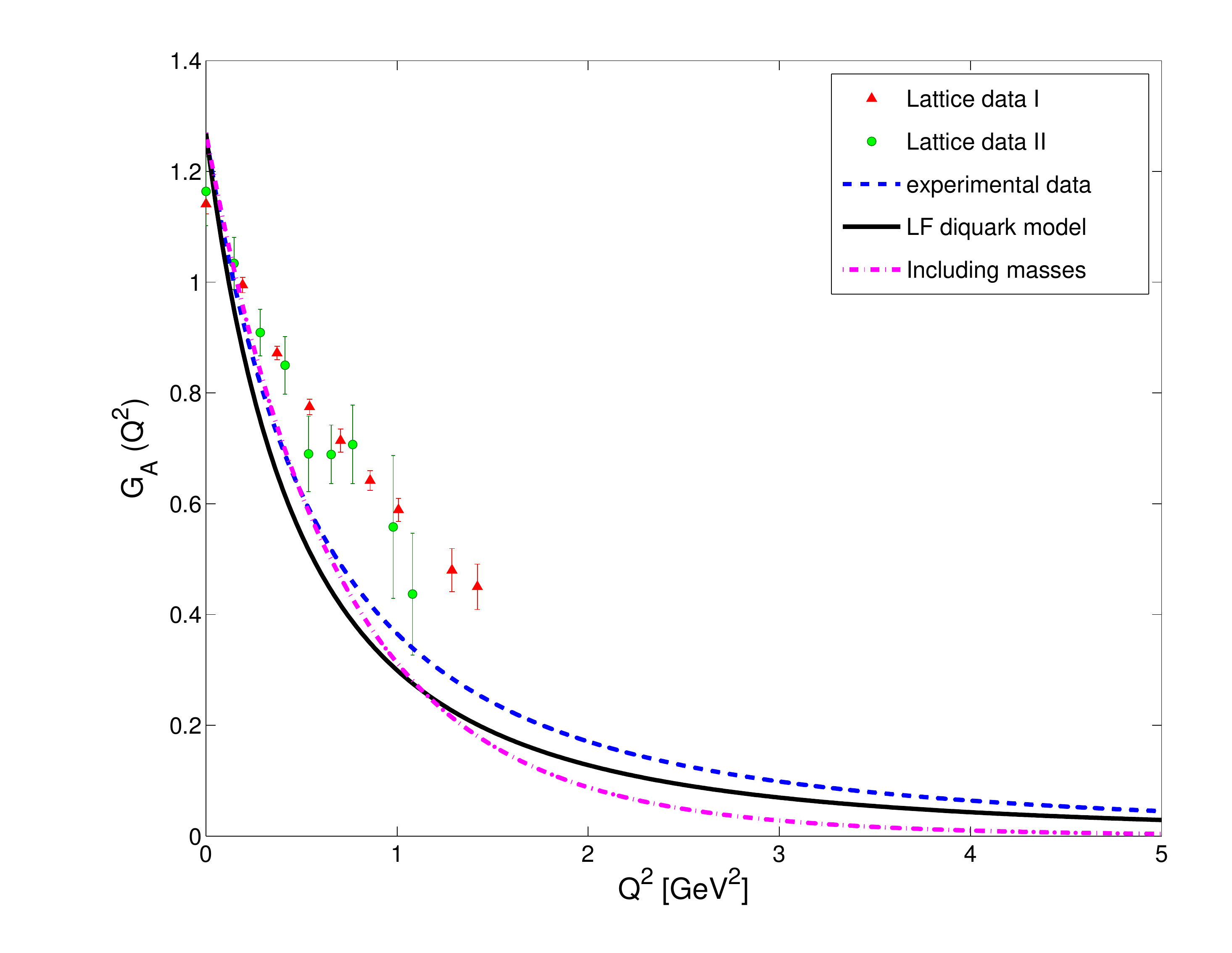}
\caption{\label{axial_FF}(Color online) Plot of the axial vector form factors $G_A(Q^2)=G_A^{u-d}(Q^2)$. The black solid line represents the quark-diquark model in AdS/QCD, the blue dashed line represents dipole fit of experimental data \cite{Bernard:2001rs} and the data are taken from lattice calculation~\cite{Alexandrou:2013joa}. The pink dashed-dot line represents the result by including of quark and diquark masses in the wavefunctions (Eq.(\ref{with_mass})). }
\end{figure}
\subsection{Mellin moments of helicity dependent GPDs}
\begin{figure*}[htbp]
\begin{minipage}[c]{0.98\textwidth}
\small{(a)}
\includegraphics[width=7.5cm,height=5.15cm,clip]{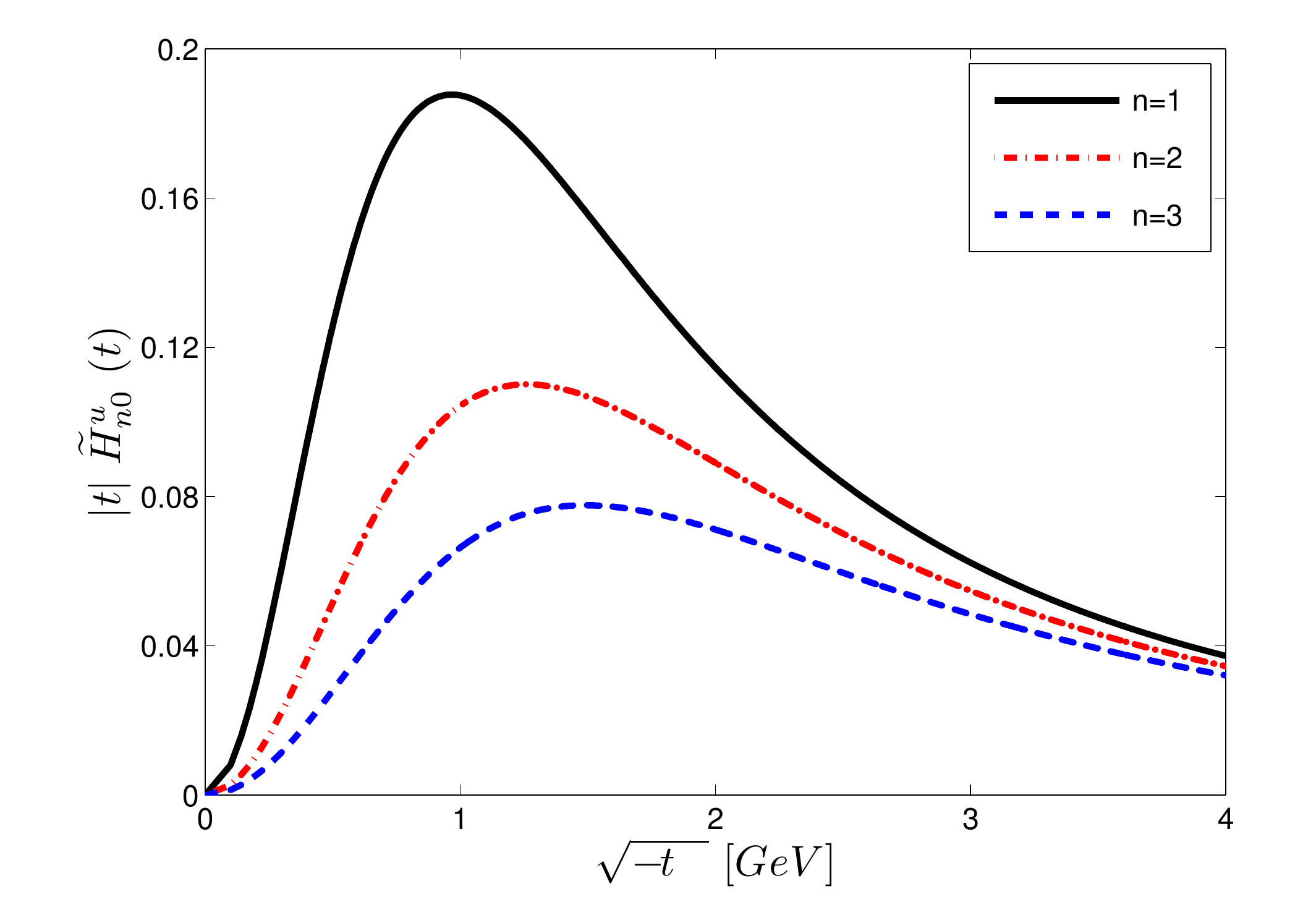}
\hspace{0.1cm}%
\small{(b)}\includegraphics[width=7.5cm,height=5.15cm,clip]{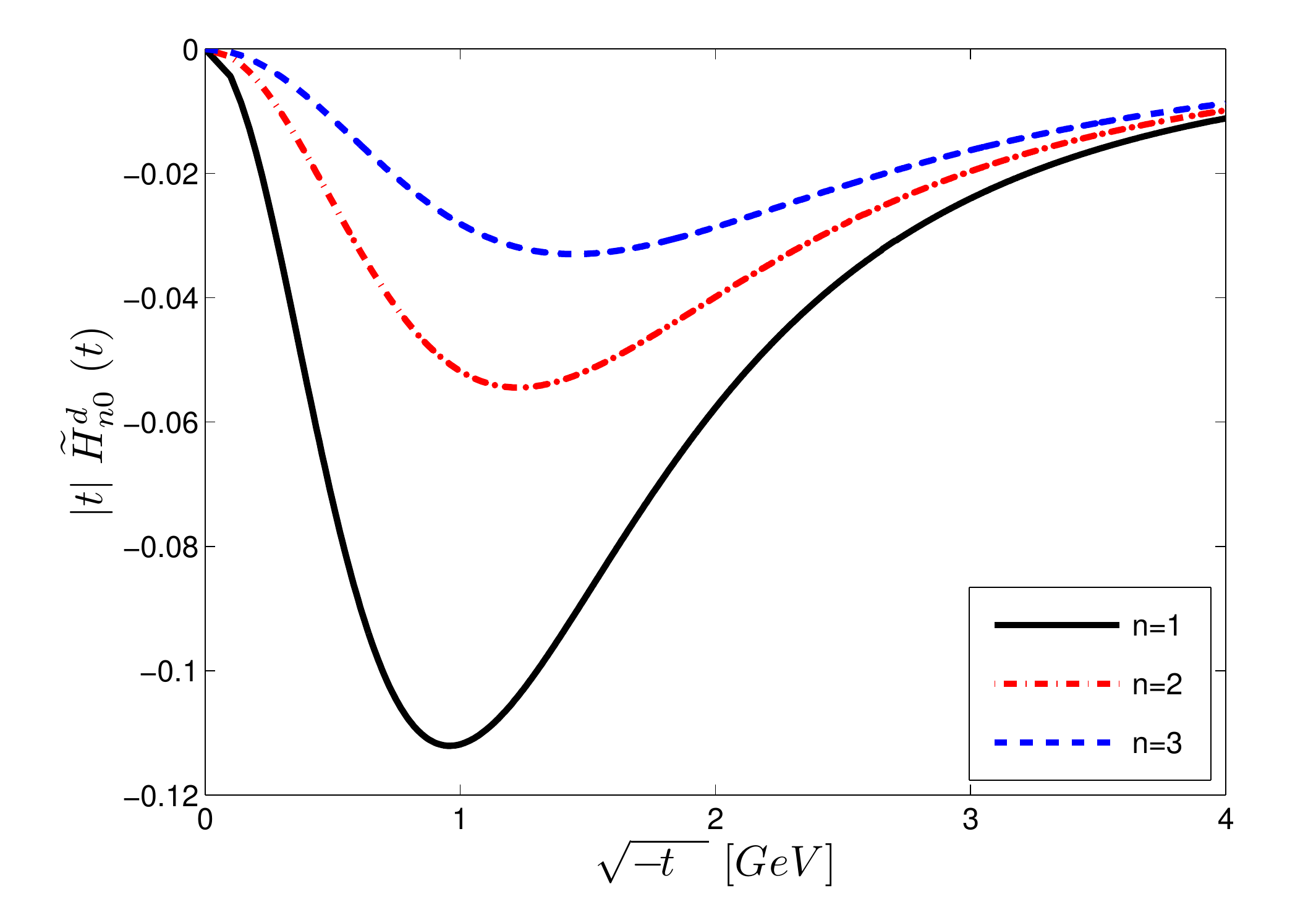}
\end{minipage}
\begin{minipage}[c]{0.98\textwidth}
\small{(c)}\includegraphics[width=7.5cm,height=5.15cm,clip]{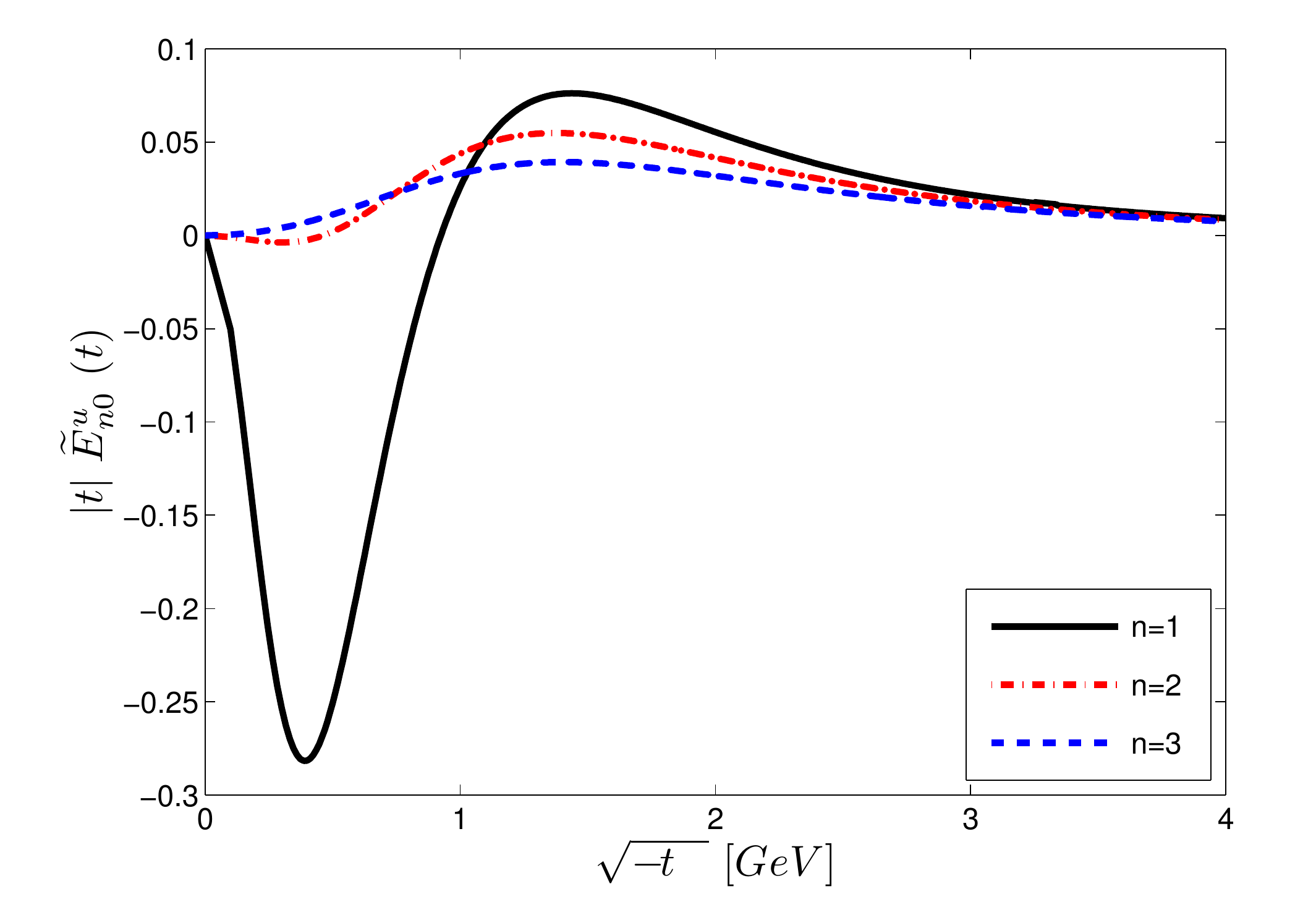}
\hspace{0.1cm}%
\small{(d)}\includegraphics[width=7.5cm,height=5.15cm,clip]{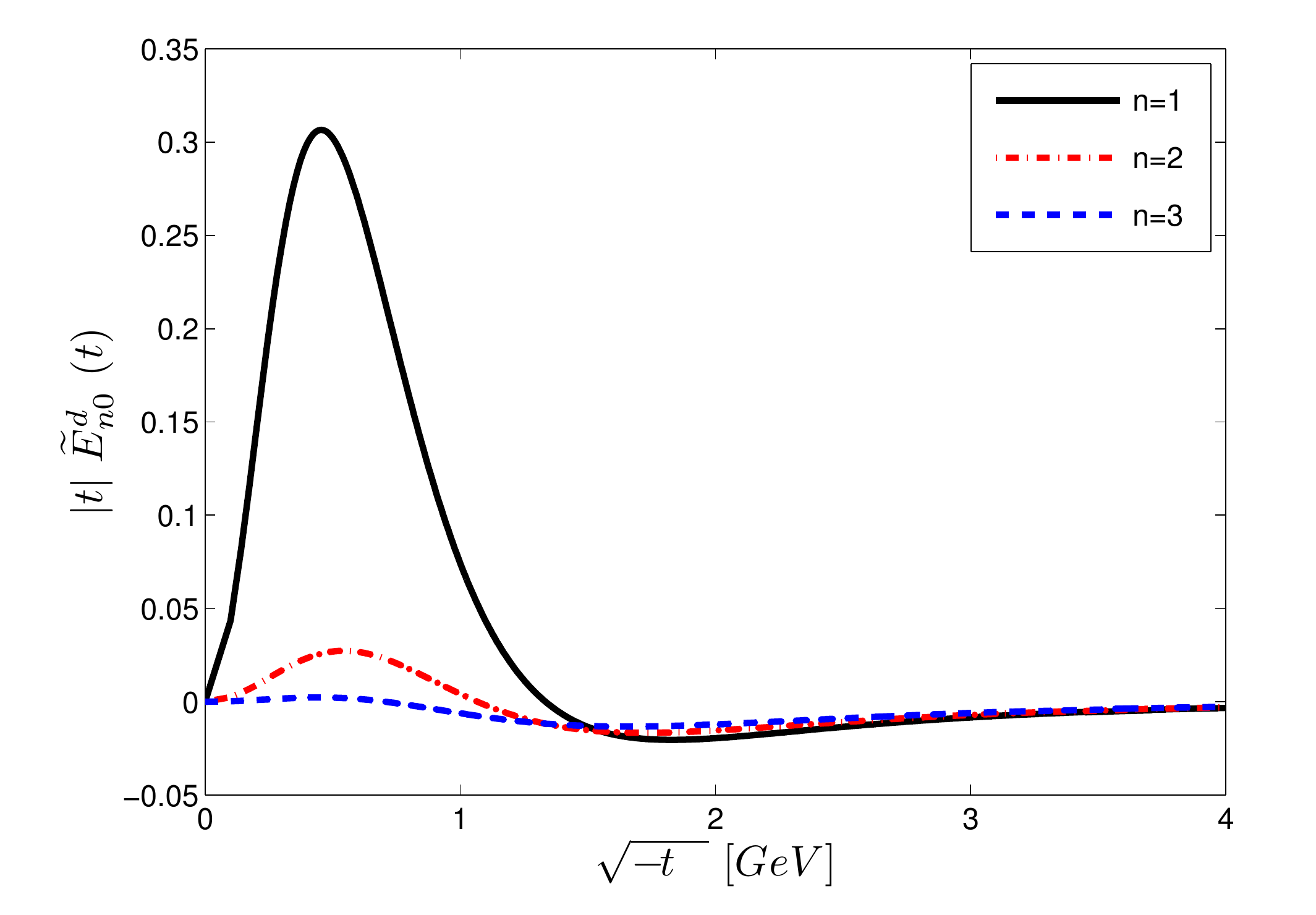}
\end{minipage}
\caption{\label{moments}(Color online) Plots of first three moments of the helicity dependent GPDs for zero skewness vs $\sqrt{-t}$ in $GeV$. Left panel is for $u$ quark and the right panel is for $d$ quark.}
\end{figure*}

The Mellin moments of the valence GPDs are defined as
\be
\widetilde{H}^q_{n0}(t)=\int_0^1~dx x^{n-1}\widetilde{H}^q(x,0,t),\label{moment_formula}
\ee
where the index $n=1,2,3$ etc., and the second subscript implies that the moments are evaluated at zero skewness. The moments of the other GPD, $\widetilde{E}^q_{n0}(t)$ are also defined in the same way as (\ref{moment_formula}).
The first moments of $\widetilde{H}^q_{n0}(x,0,t)$ and  $\widetilde{E}^q_{n0}(x,0,t)$ give the axial vector form factor, $G_A^q(t)$ and pseudoscalar form factor, $G_p^q(t)$ for quark $q$, respectively. The forward value, $t = 0$, of the form factors $g_A=\widetilde{H}_{10}(t=0)$ can be identified as the axial-vector coupling constant (axial charge)~\cite{Bernard:2001rs,Hagler:2009}. Similarly, $g_P=\widetilde{E}_{10}(t=0)$ is known as the pseudo-scalar coupling constant. In Fig.\ref{axial_FF}, we compare the result for the axial vector form factors obtained in the quark-diquark model in AdS/QCD with the corresponding results from lattice~\cite{Alexandrou:2013joa} and the experimental data described by the dipole formula \cite{Bernard:2001rs}:
\be
G_A(Q^2)=\frac{g_A}{(1+Q^2/M_A^2)^2}
\ee
where the axial-vector coupling constant, $g_A=1.2673$ and the the parameter $M_A=1.069$ GeV, the so-called axial mass \cite{Bernard:2001rs}. The plot shows that our result is more or less in agreement with the dipole fit of experimental data. In the same plot,
we also compare the result of axial form factor by introducing the mass terms in the wavefunctions $\varphi_q^{(i)}(x,\bfk)$ (Eq.(\ref{wf2})), following the Ref. \cite{Brodsky:2008pg}
\be\label{with_mass}
&&\varphi_q^{(i)}(x,\bfk)\nonumber\\&\sim&\exp\bigg[-\frac{\bfk^2}{2\kappa^2}\Big\{\frac{\log(1/x)}{(1-x)^2}+\frac{m_q^2}{x}+\frac{m_D^2}{(1-x)}\Big\}\bigg].
\ee
Here we use the quark and diquark masses $m_q=0.35$ GeV and $m_D=0.65$ GeV respectively. With the mass terms in the wavefunctions, the result is in good agreement with the experimental data at low $Q^2$, however, it deviates at higher $Q^2$.
The second moments of these GPDs correspond to the gravitational form factors of longitudinally polarized quarks in an unpolarized nucleon. The third moments of the GPDs give form factors of a twist-two operator having
two covariant derivatives~\cite{rev} and the higher order moments generate the form factors of higher-twist operators. 
In Fig.\ref{moments}, the first three moments of the helicity dependent GPDs $|t|\widetilde{H}^q_{n0}(t)$, $|t|\widetilde{E}^q_{n0}(t)$ as functions of $\sqrt{-t}$ have been shown for $u$ and $d$ quarks. We observe a strong decrease in the  magnitudes of the moments with increasing  $n$. One can understand this aspect from the behavior of the GPDs with $x$ as shown in Fig.\ref{gt7}. Since higher moments involve higher power of $x$, the dominant contributions appears from the large $x$ region($x\to1$). But the GPDs decrease rapidly as $x$ increases, thus the higher moments become smaller.
One can also observe that with increasing the index $n$, the decrease of the moments becomes slower as $-t$ increases. This phenomena again can be  described in terms of the decrease of the GPDs with momentum fraction $x$, which shows in a weaker $t$ slope for the higher moments. A similar behavior of the GPDs has been in others phenomenological models \cite{Diehl:2004cx,Chakrabarti:2015ama,neetika} and in lattice QCD ~\cite{Hagler:2007xi,Hagler:2009,gock}.  

\section{Impact parameter representation of helicity dependent GPDs}\label{helicity_impact}
\subsection{GPDs in transverse impact parameter space}\label{helicity_trans_impact}
The transverse impact parameter dependent GPDs are defined by a two-dimensional Fourier transform with respect to the momentum transfer in the transverse direction ~\cite{burk1,burk2,Diehl:2002he}:
\be
\widetilde{H}^q(x,\zeta,{b})&=&{1\over (2 \pi)^2} \int d^2 {\bf \Delta}_{\perp} e^{-i {\bf \Delta}_{\perp} \cdot {\bf b}_{\perp}}
\widetilde{H}^q(x,\zeta,t),\\
\widetilde{E}^q(x,\zeta,{b})&=&{1\over (2 \pi)^2} \int d^2 {\bf \Delta}_{\perp} e^{-i {\bf \Delta}_{\perp} \cdot {\bf b}_{\perp}}
\widetilde{E}^q(x,\zeta,t),
\ee
where ${\bf b}_{\perp}$ represents the transverse impact parameter conjugate to the transverse momentum transfer ${\bf \Delta}_{\perp}$. 
For zero skewness, $b=|\bfb|$ corresponds a measure of  the transverse distance of the struck parton from the center of momentum of the hadron and it follows the condition $\sum_i x_i b_{i}=0$, where the sum is over the number of partons. The relative distance between the struck parton and the center of momentum of the spectator system is given by  ${\mid \bfb \mid\over 1-x}$, which provides us an estimate of the size of the bound state \cite{diehl}. For nonzero $\zeta$, the transverse distance of partons from the proton center of momentum differs in the initial and final state, but their relative distance to each other in a hadron stays the same. The transverse position $\bfb$, with the initial and final state proton is shifted relative to each other by an amount of order $\zeta \bfb$ \cite{Diehl:2002he}.
In the DGLAP region $x>\zeta$, the impact parameter ${\bf b}_{\perp}$ describes the location where the quark is pulled out and re-insert to the proton. In the ERBL domain $x<\zeta$, ${\bf b}_{\perp}$ gives the transverse distance of the quark-antiquark pair inside the proton. For zero skewness, the helicity dependent GPDs also have a density interpretation in transverse impact parameter space like other GPDs corresponding the density for longitudinally polarized partons. $\widetilde{H}^q(x,b)$ reflects the density of quarks with helicity equal or opposite to the proton helicity ~\cite{Boffi:2007yc,Diehl05,Pasquini2}.  Note that the density interpretation is possible only in the limit $\zeta=0$, but it is natural to ask how this
situation looks like at nonzero $\zeta$, which is applicable for most processes where GPDs
can be accessed. Thus, it is interesting to study the helicity dependent GPDs in the impact parameter space when $\zeta$ is nonzero. 
\begin{figure*}[htbp]
\begin{minipage}[c]{0.98\textwidth}
\small{(a)}
\includegraphics[width=7.5cm,height=5.15cm,clip]{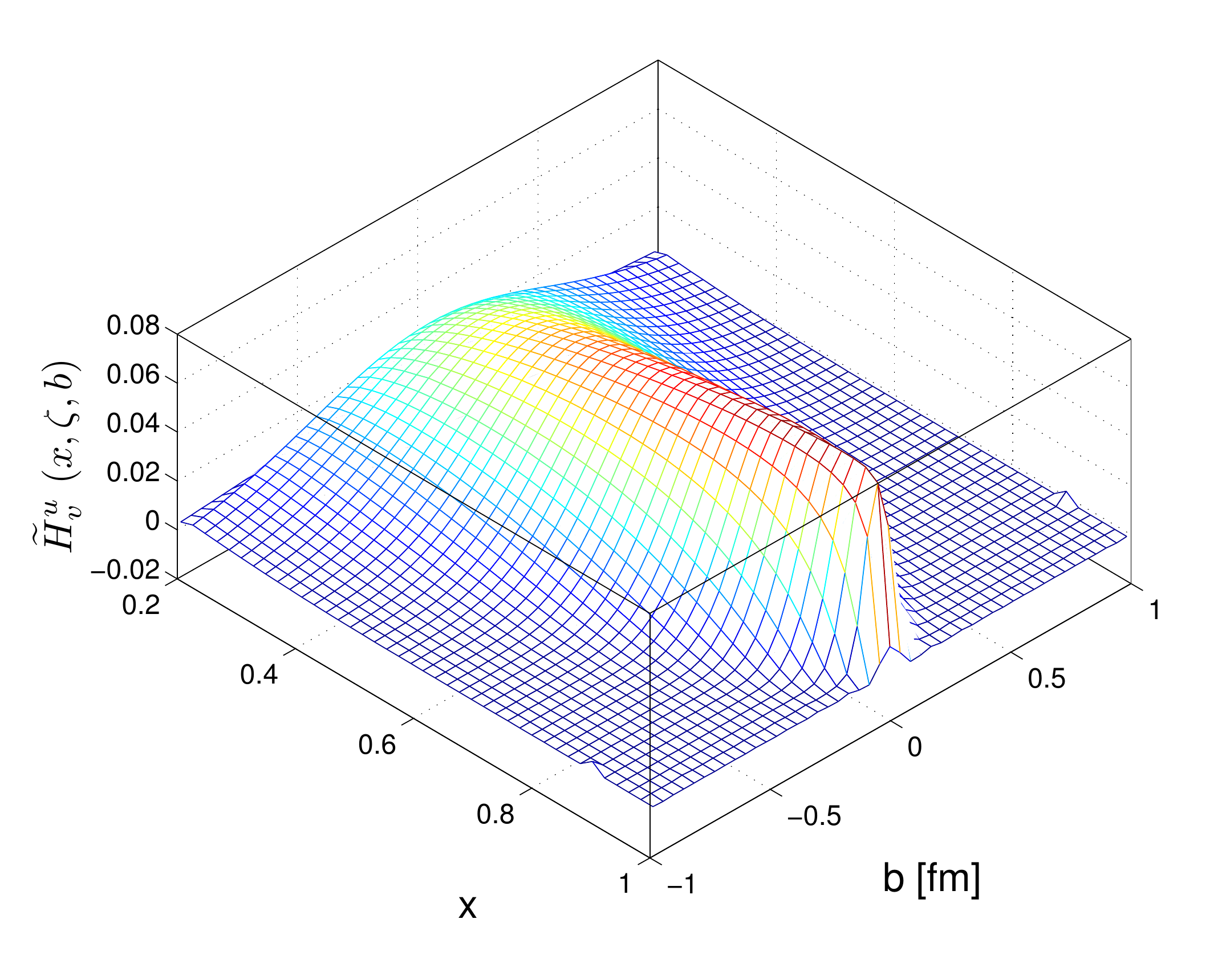}
\hspace{0.1cm}%
\small{(b)}\includegraphics[width=7.5cm,height=5.15cm,clip]{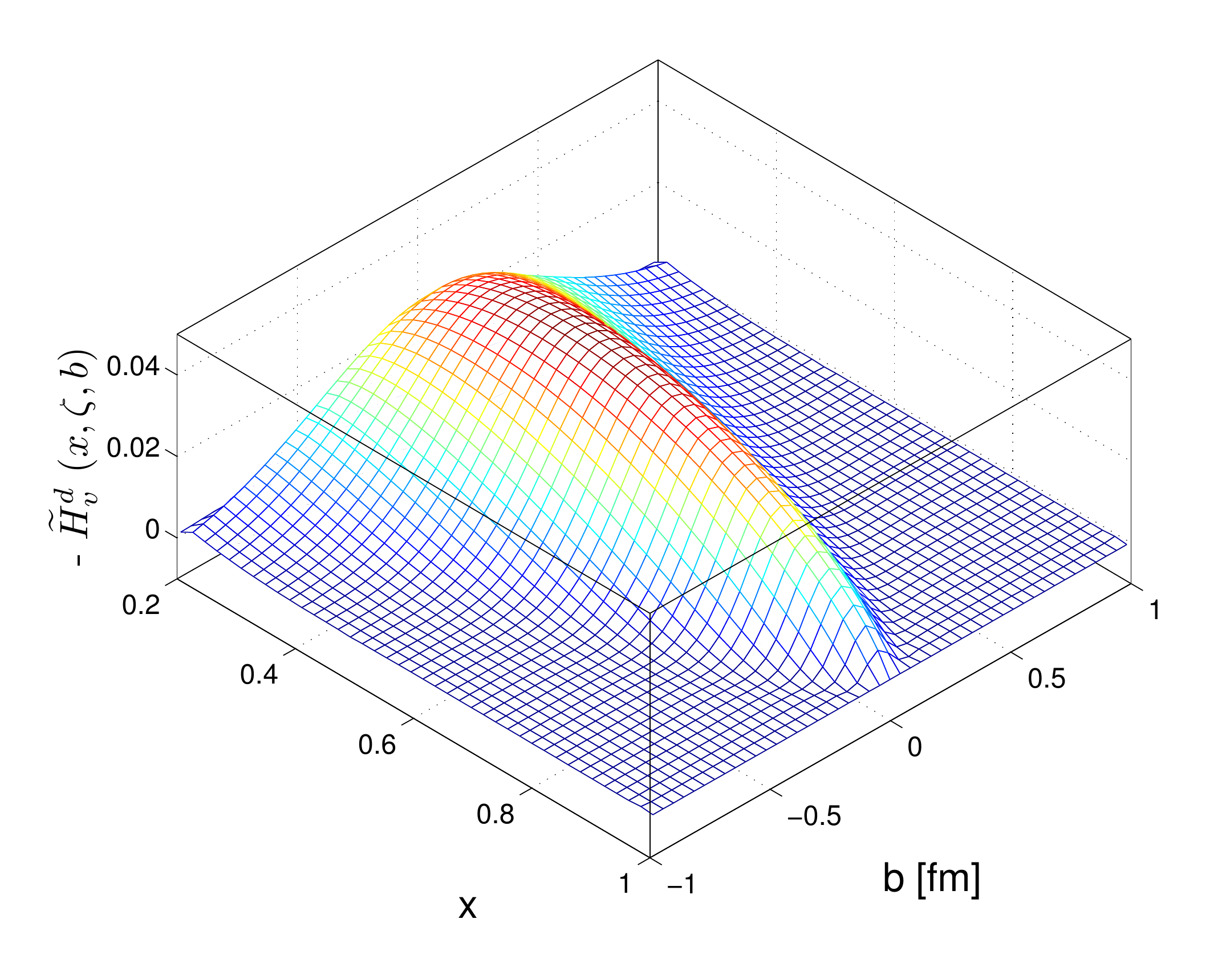}
\end{minipage}
\begin{minipage}[c]{0.98\textwidth}
\small{(c)}\includegraphics[width=7.5cm,height=5.15cm,clip]{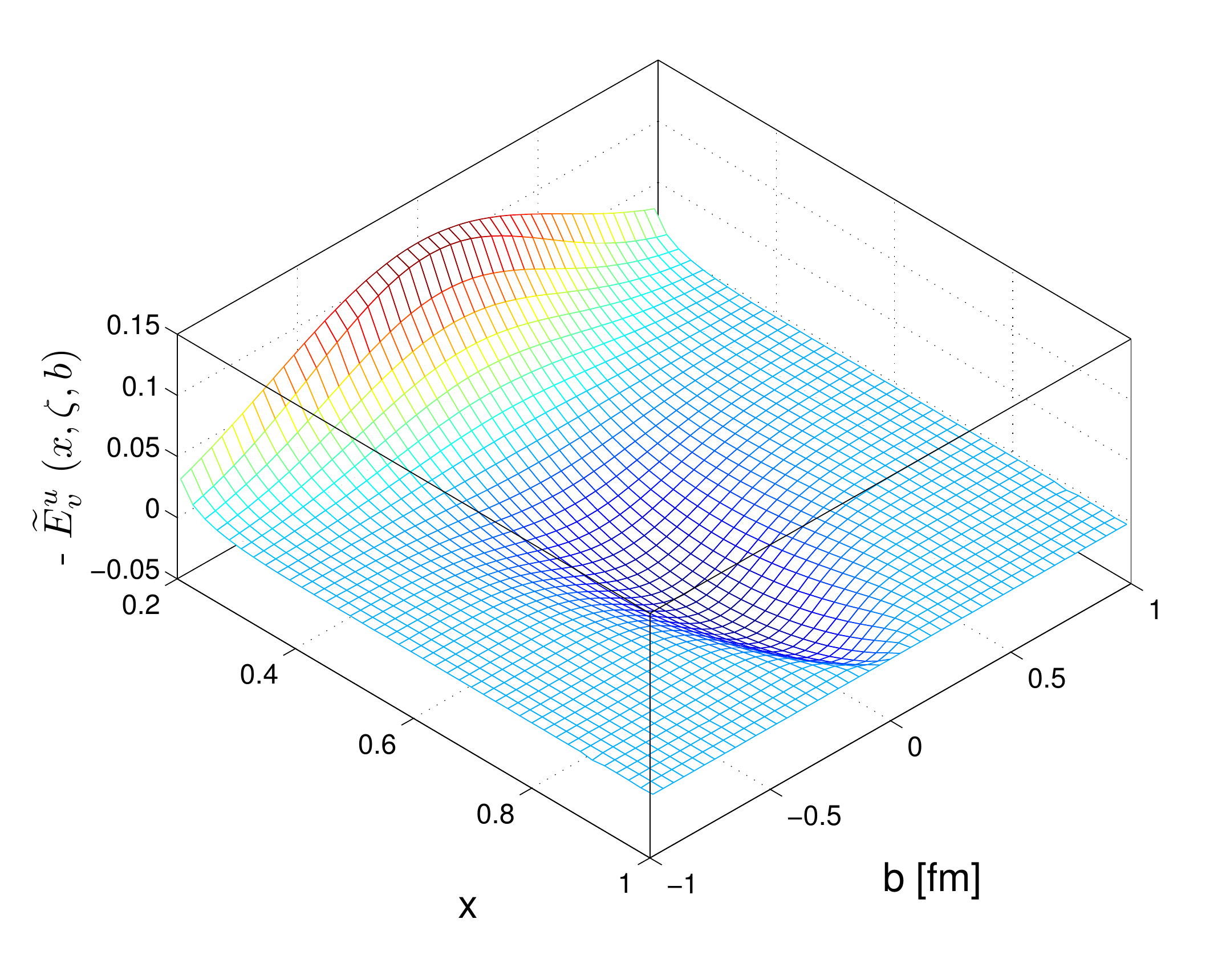}
\hspace{0.1cm}%
\small{(d)}\includegraphics[width=7.5cm,height=5.15cm,clip]{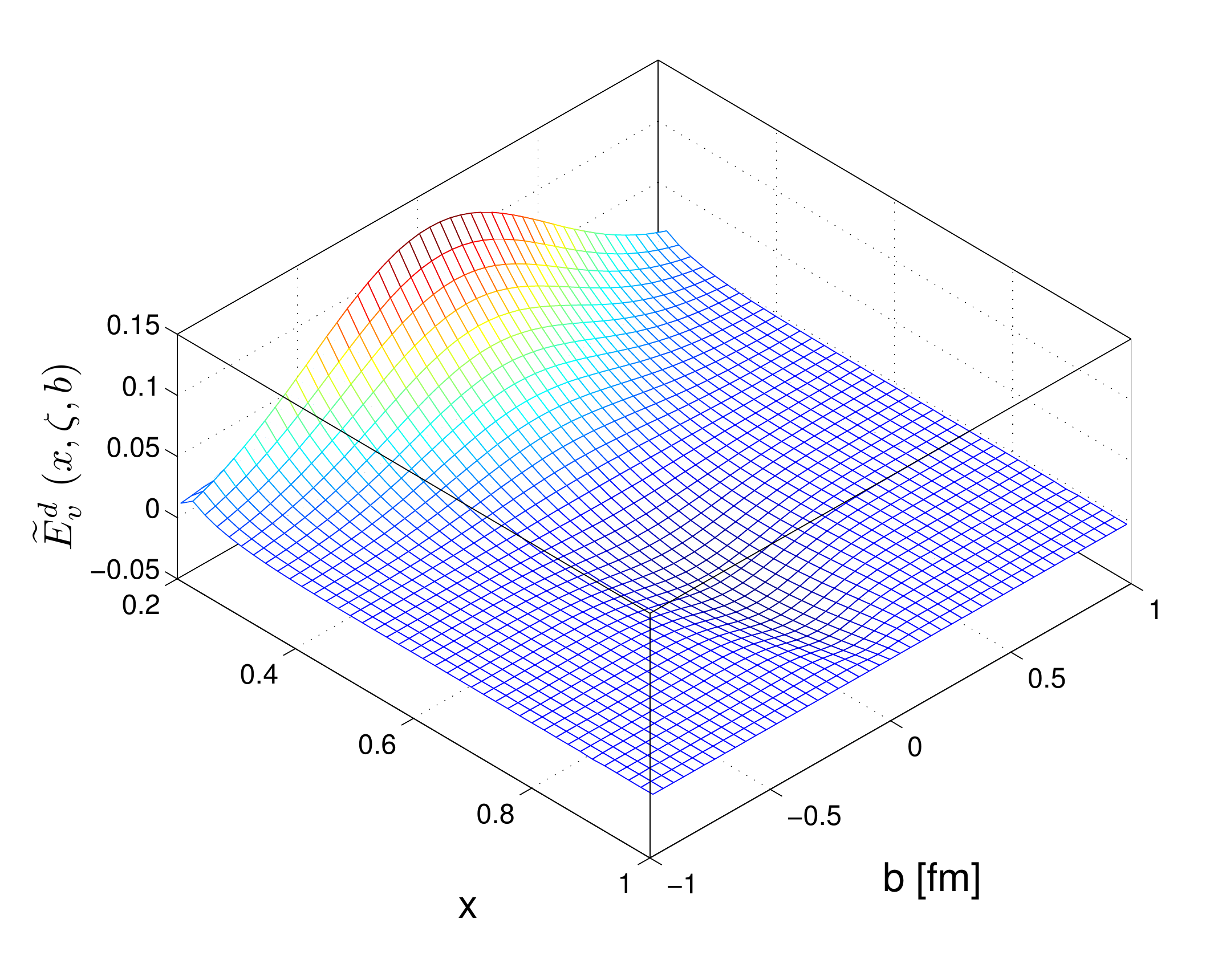}
\end{minipage}
\caption{\label{giz2}(Color online) Plots of helicity dependent GPDs for the nonzero skewness in impact space vs $x$ and $b=|\bf b|$ for fixed value of $\zeta=0.2$. Left panel is for $u$ quark and the right panel is for $d$ quark.}
\end{figure*}
\begin{figure*}[htbp]
\begin{minipage}[c]{0.98\textwidth}
\small{(a)}
\includegraphics[width=7.5cm,height=5.15cm,clip]{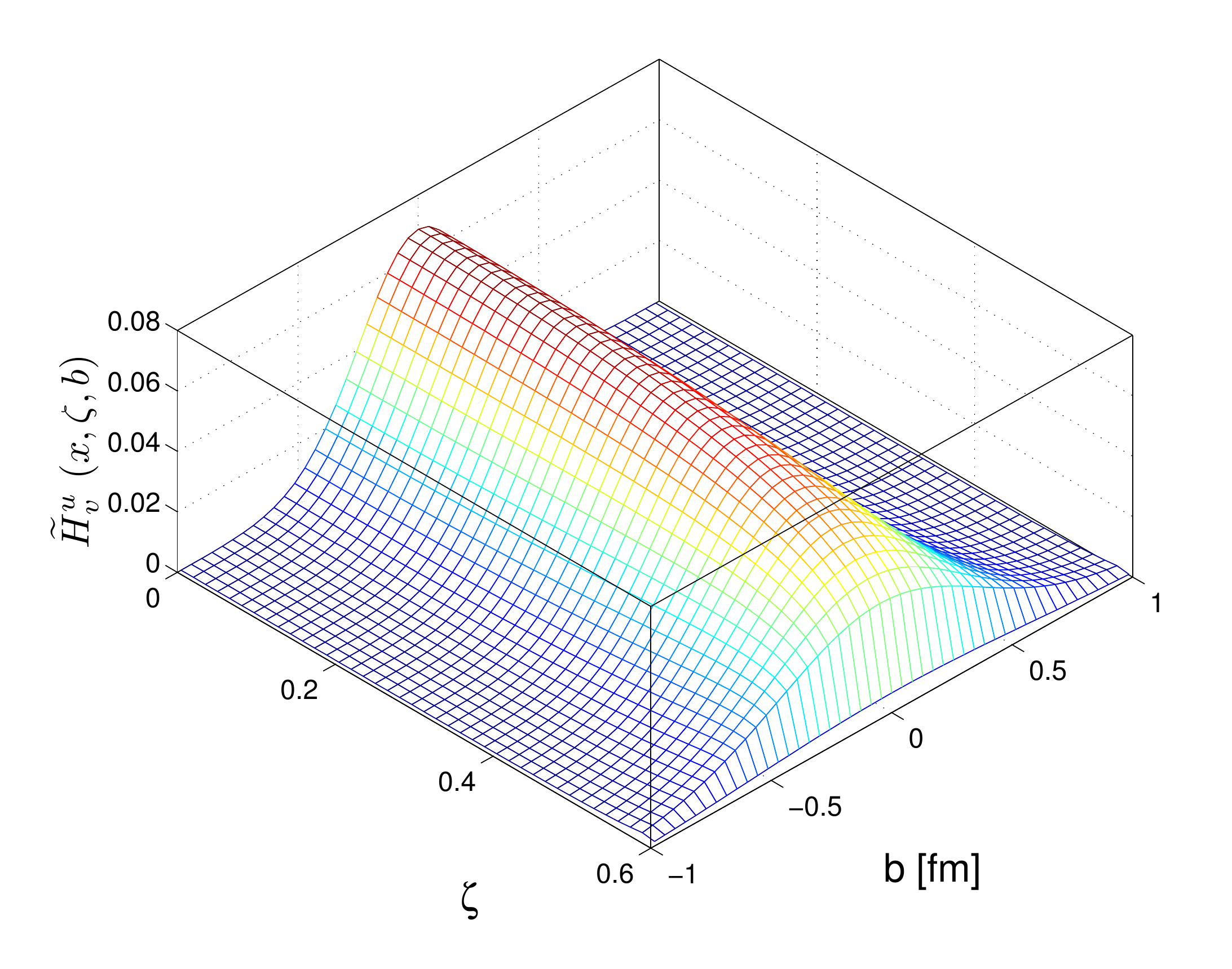}
\hspace{0.1cm}%
\small{(b)}\includegraphics[width=7.5cm,height=5.15cm,clip]{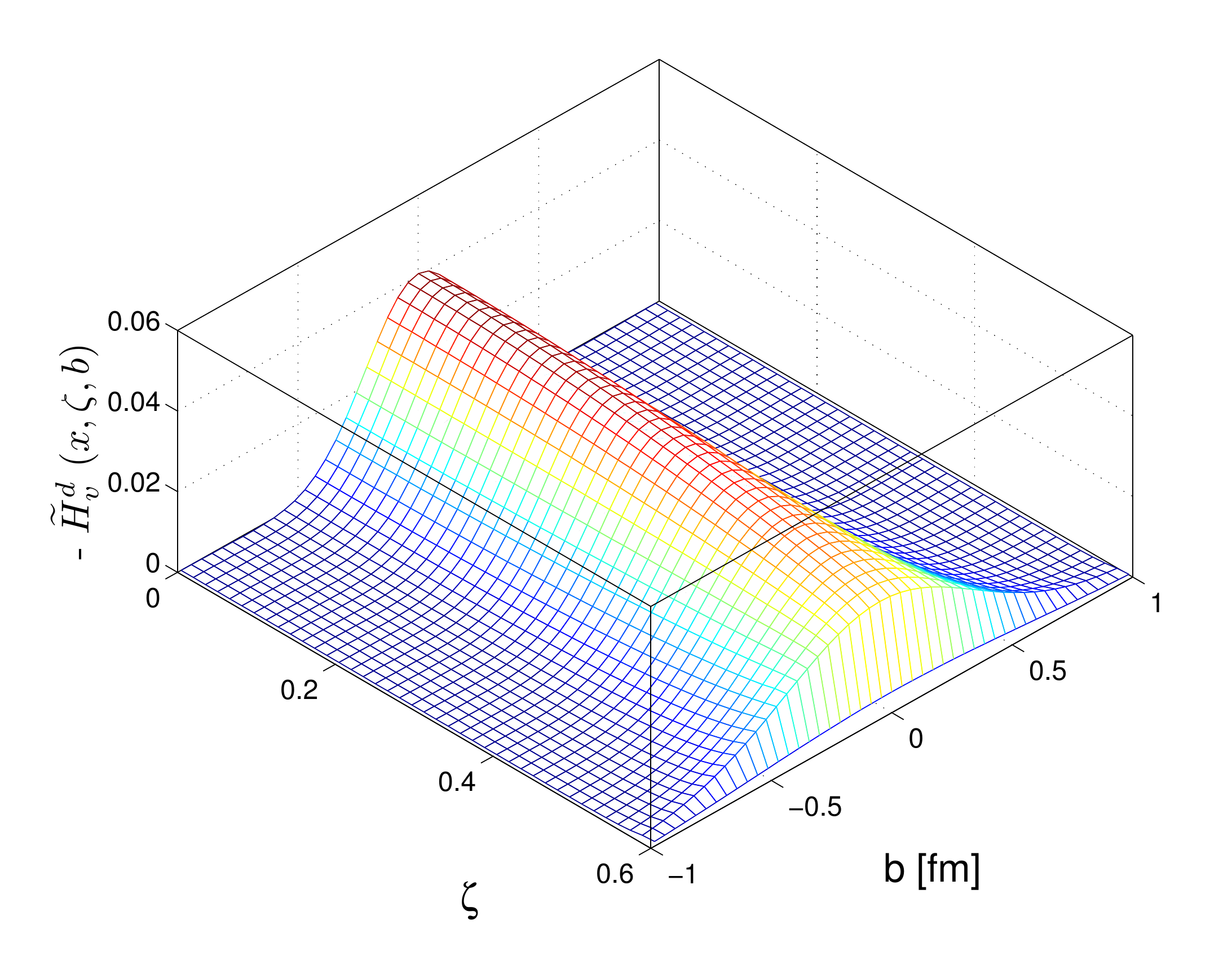}
\end{minipage}
\begin{minipage}[c]{0.98\textwidth}
\small{(c)}\includegraphics[width=7.5cm,height=5.15cm,clip]{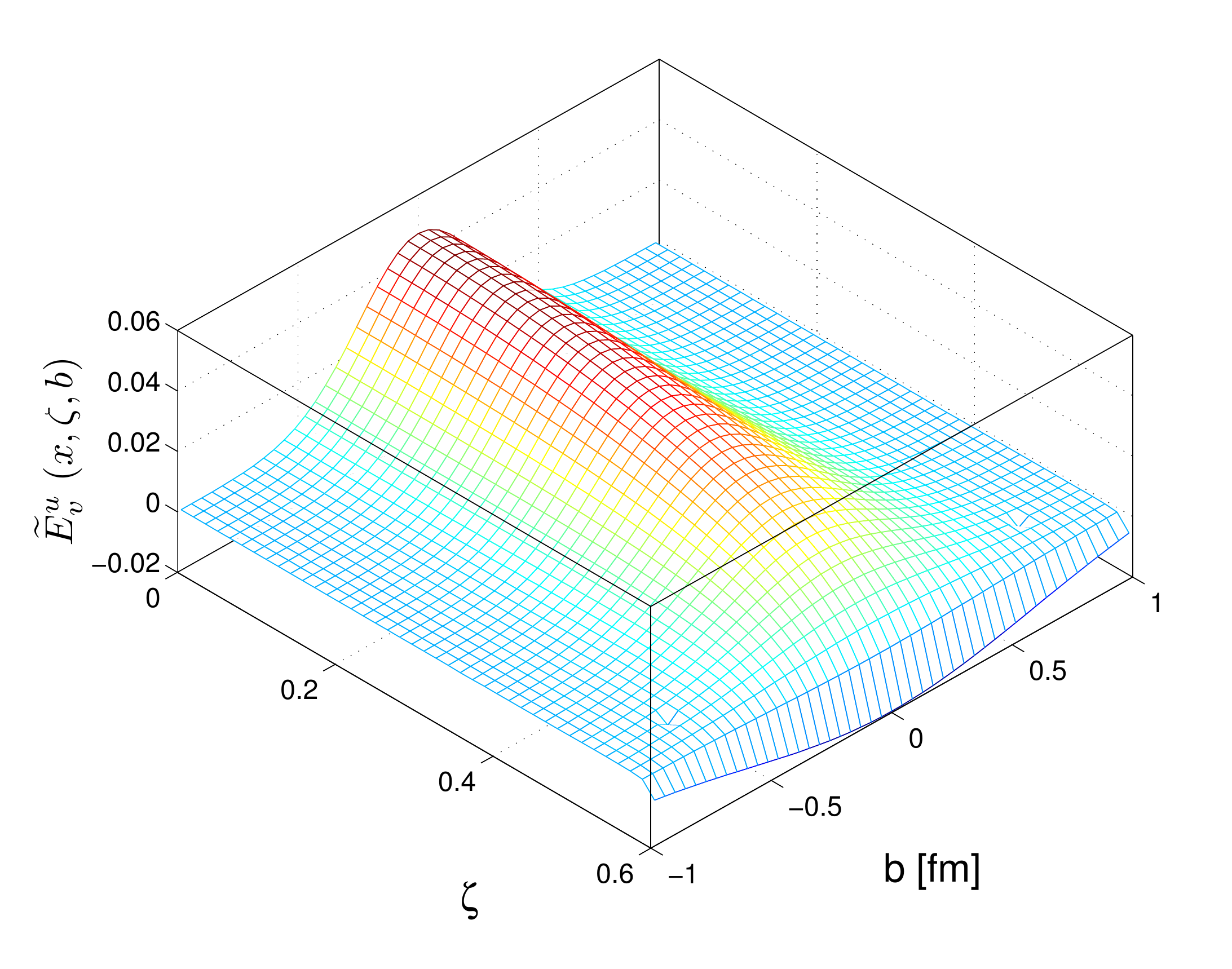}
\hspace{0.1cm}%
\small{(d)}\includegraphics[width=7.5cm,height=5.15cm,clip]{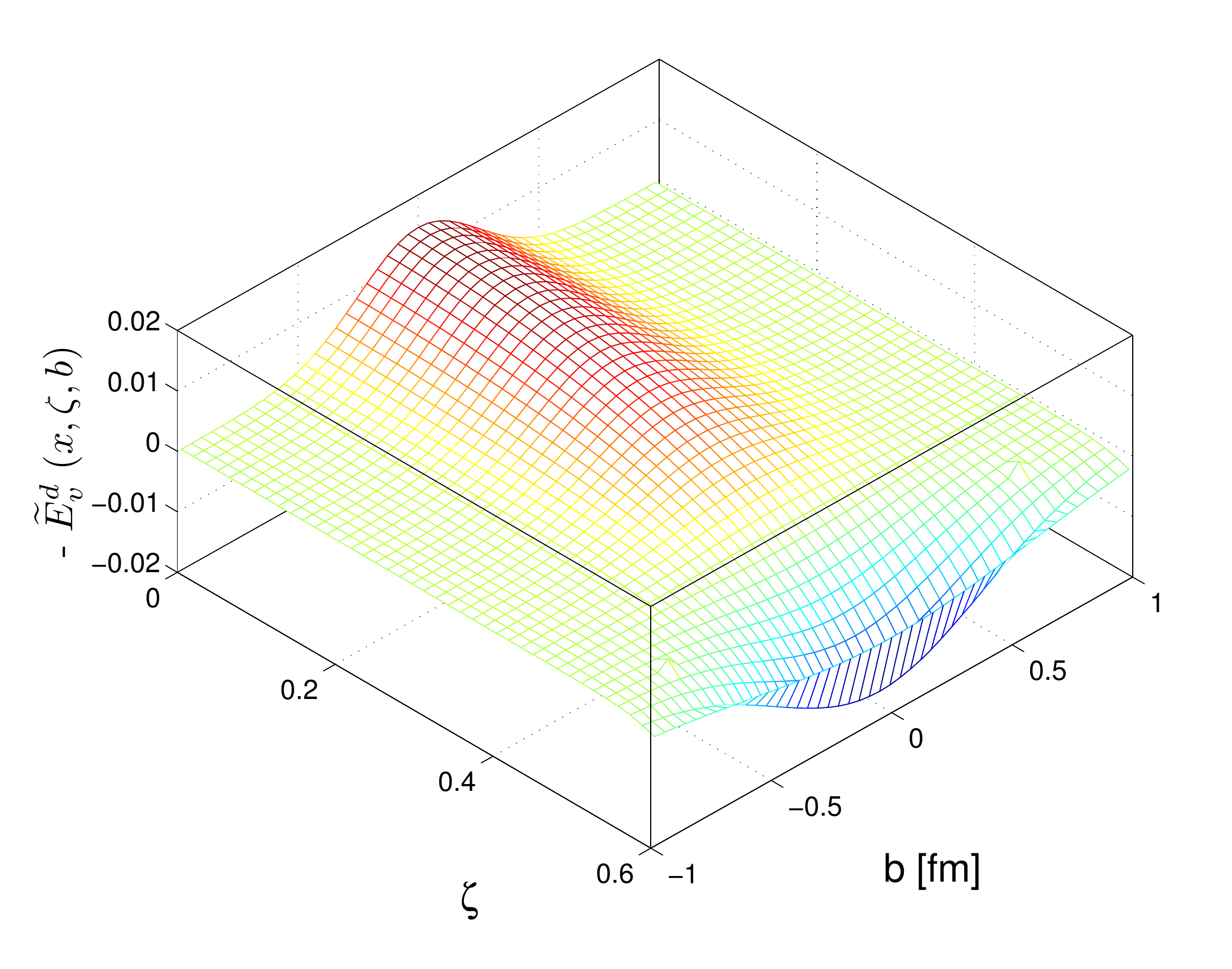}
\end{minipage}
\caption{\label{gix6}(Color online) Plots of helicity dependent GPDs for the nonzero skewness in impact space vs $\zeta$ and $b=|\bf b|$ for fixed value of $x=0.6$. Left panel is for $u$ quark and the right panel is for $d$ quark.}
\end{figure*}

In Fig.\ref{giz2}, we show the skewness dependent GPDs $\widetilde{H}(x,\zeta,b)$ and $\widetilde{E}(x,\zeta,b)$ for $u$ and $d$ quark in transverse impact parameter space for fixed $\zeta=0.2$ as functions of $b$ and $x$. Similarly, the GPDs as functions of $\zeta$ and $b$ for a fixed value of $x=0.6$ are shown in Fig.\ref{gix6}. 
The peak of the distribution $\widetilde{H}(x,\zeta,b)$ for fixed $\zeta$ appears at higher $x$ for $u$ quark whereas it shifts to lower $x$ for $d$ quark. $\widetilde{E}(x,\zeta,b)$ shows the peaks at lower $x$ for both $u$ and $d$ quarks and one can also observe an oscillatory behavior for the GPDs, $\widetilde{E}(x,\zeta,b)$. This is due to the fact that the GPD in momentum space, $\widetilde{E}(x,\zeta,t)$  has slight oscillatory behavior as can be seen in  Fig.\ref{gt7}(b). 
The width of all the distributions in transverse impact parameter space decreases with increasing $x$. This implies that the distributions are more localized near the center of momentum for higher values of $x$. We observe a similar behavior for $u$ and $d$ quark in $\widetilde{H}(x,\zeta,{b})$ and $\widetilde{E}(x,\zeta,{b})$ when they are plotted against $\zeta$ and $b$ for fixed values of $x$ in Fig.\ref{gix6}. 
Another interesting behavior of the GPDs is that for a fixed value of $x$, as $\zeta$ increases  the peaks of all the distributions become broader.  This means that as the  momentum transfer in the longitudinal direction increases the transverse distance of the longitudinally polarized active quark increases. This is due to the fact that for nonzero $\zeta$, the relative transverse distance $b$ is shifted by an amount of order $\zeta b$ \cite{Diehl:2002he}. A Similar behavior has also been observed in other phenomenological model \cite{CMM2}. We should mention here that the unpolarized, as well as the chiral-odd GPDs also exhibit a similar behavior \cite{Mondal:2015uha,Chakrabarti:2015ama}, thus one can conclude that this phenomenon of the GPDs is independent of quark polarization.   

\subsection{GPDs in longitudinal impact parameter space}\label{helicity_longi_impact}
\begin{figure*}[htbp]
\begin{minipage}[c]{0.98\textwidth}
\small{(a)}
\includegraphics[width=7.5cm,height=5.15cm,clip]{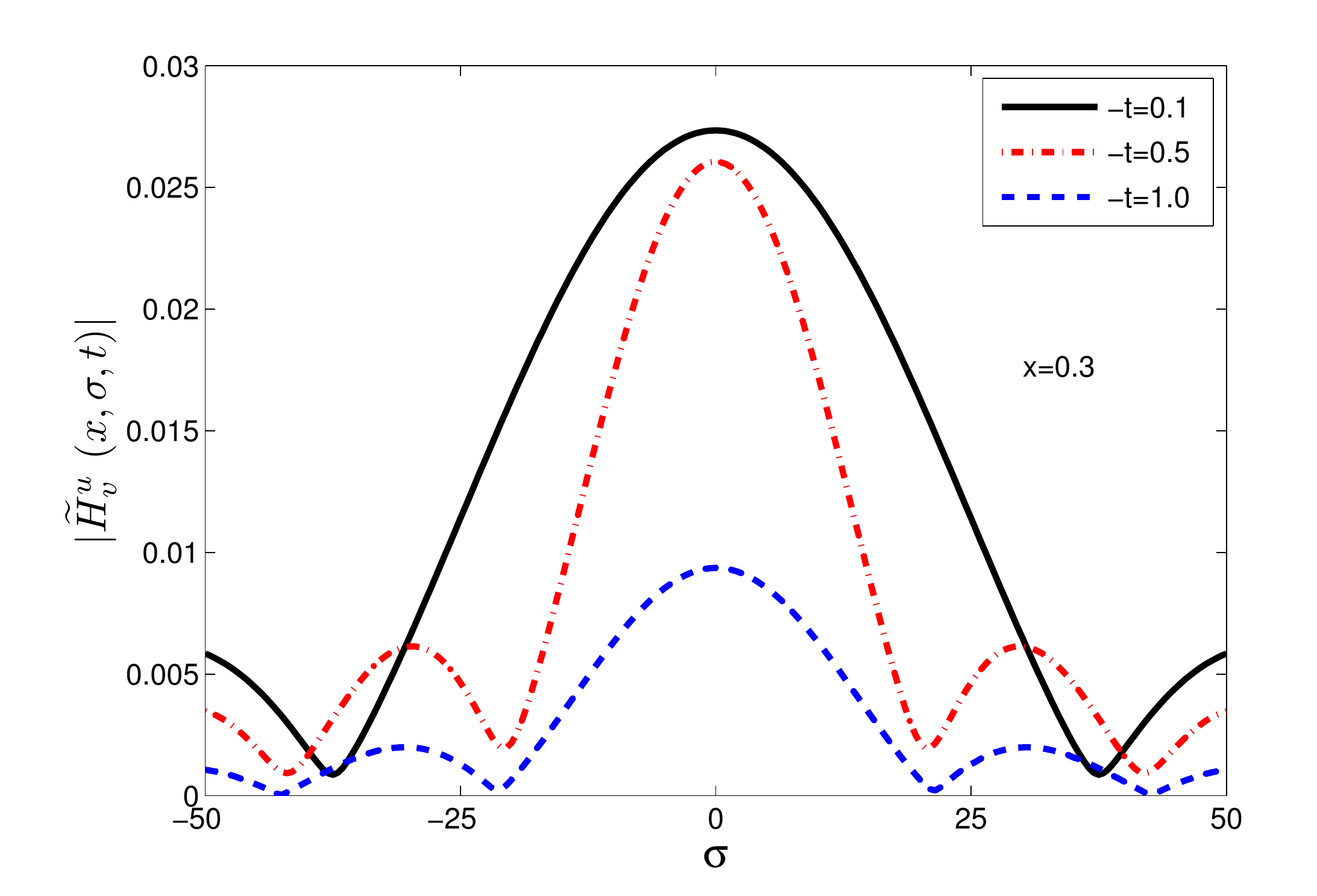}
\hspace{0.1cm}%
\small{(b)}\includegraphics[width=7.5cm,height=5.15cm,clip]{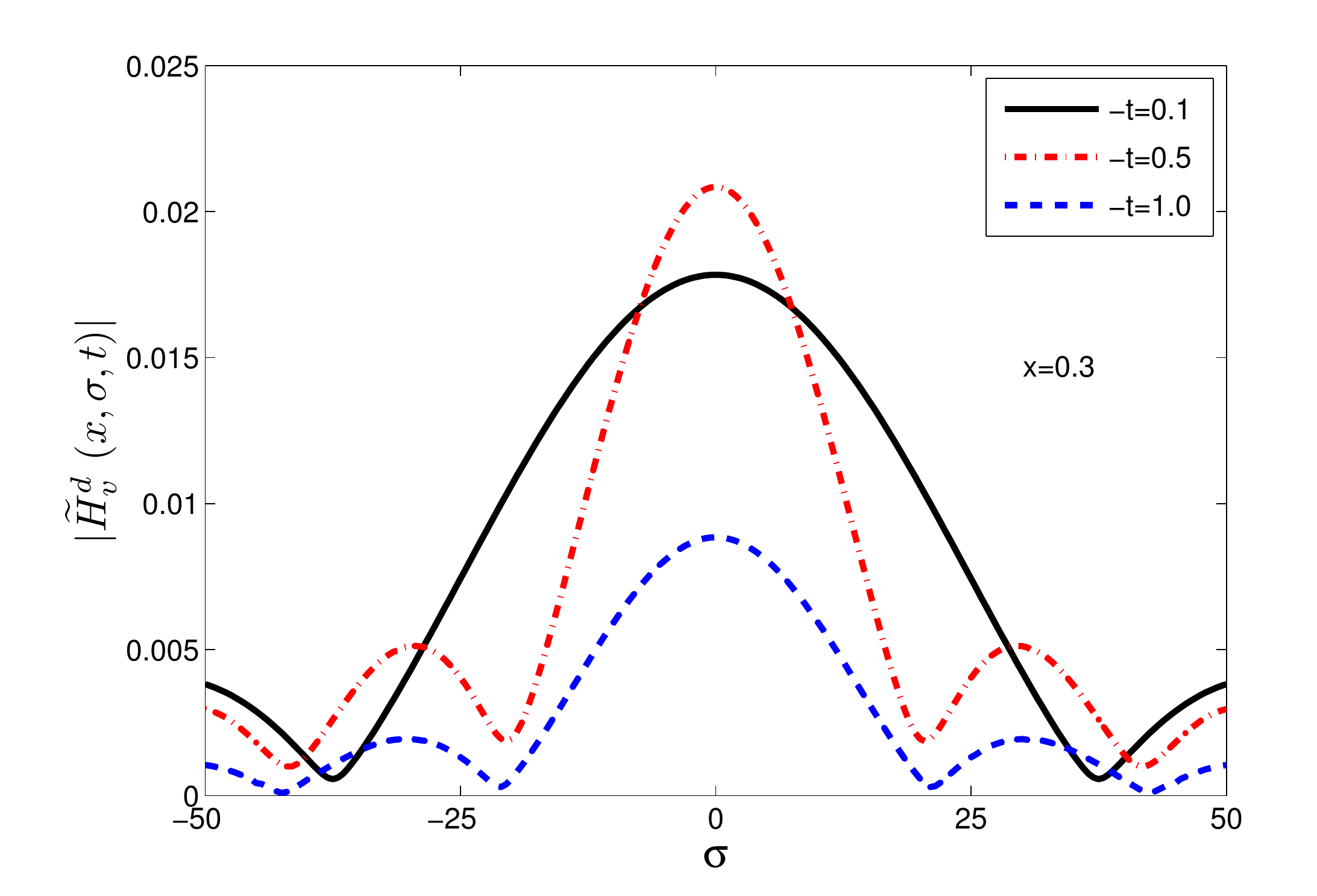}
\end{minipage}
\begin{minipage}[c]{0.98\textwidth}
\small{(c)}\includegraphics[width=7.5cm,height=5.15cm,clip]{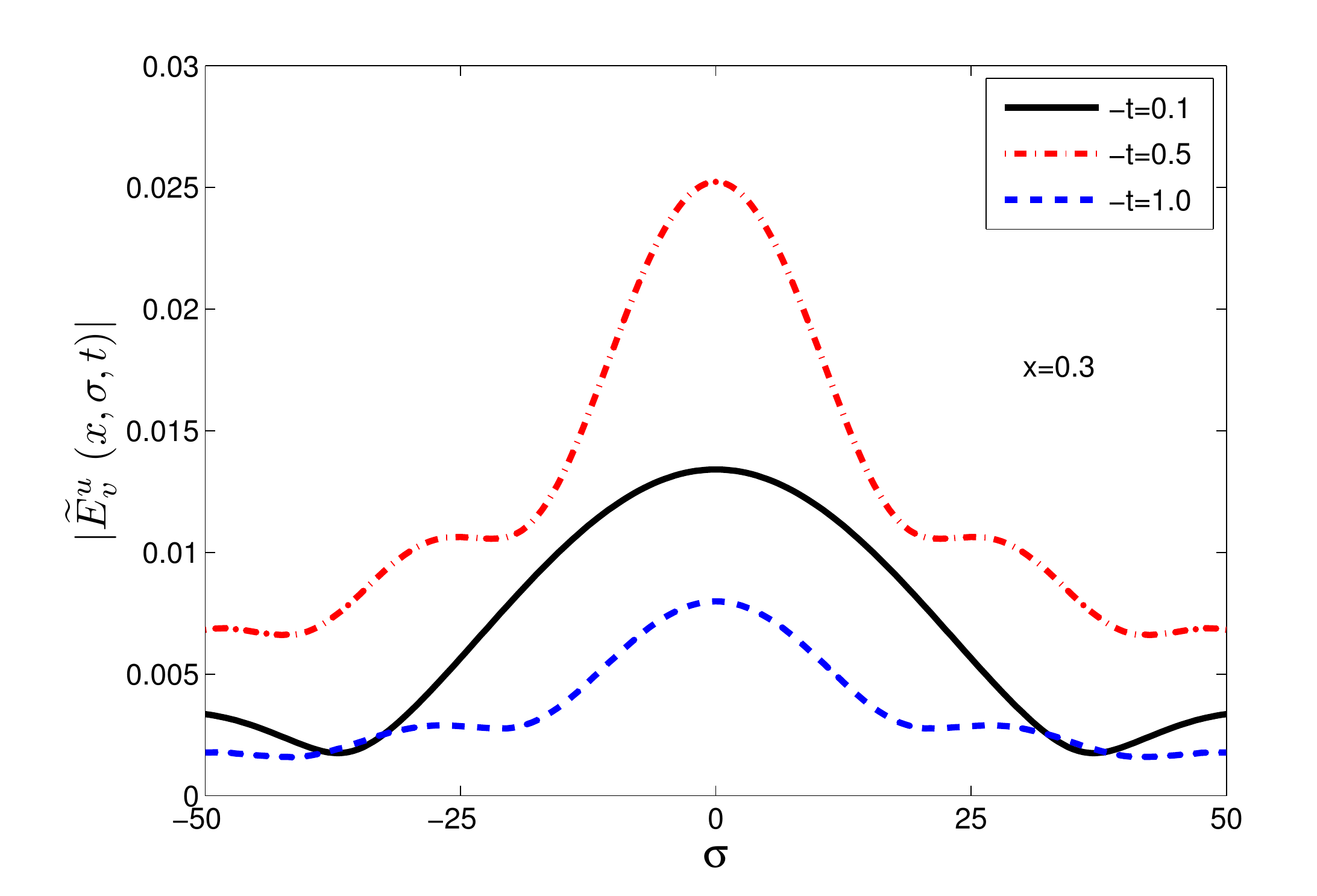}
\hspace{0.1cm}%
\small{(d)}\includegraphics[width=7.5cm,height=5.15cm,clip]{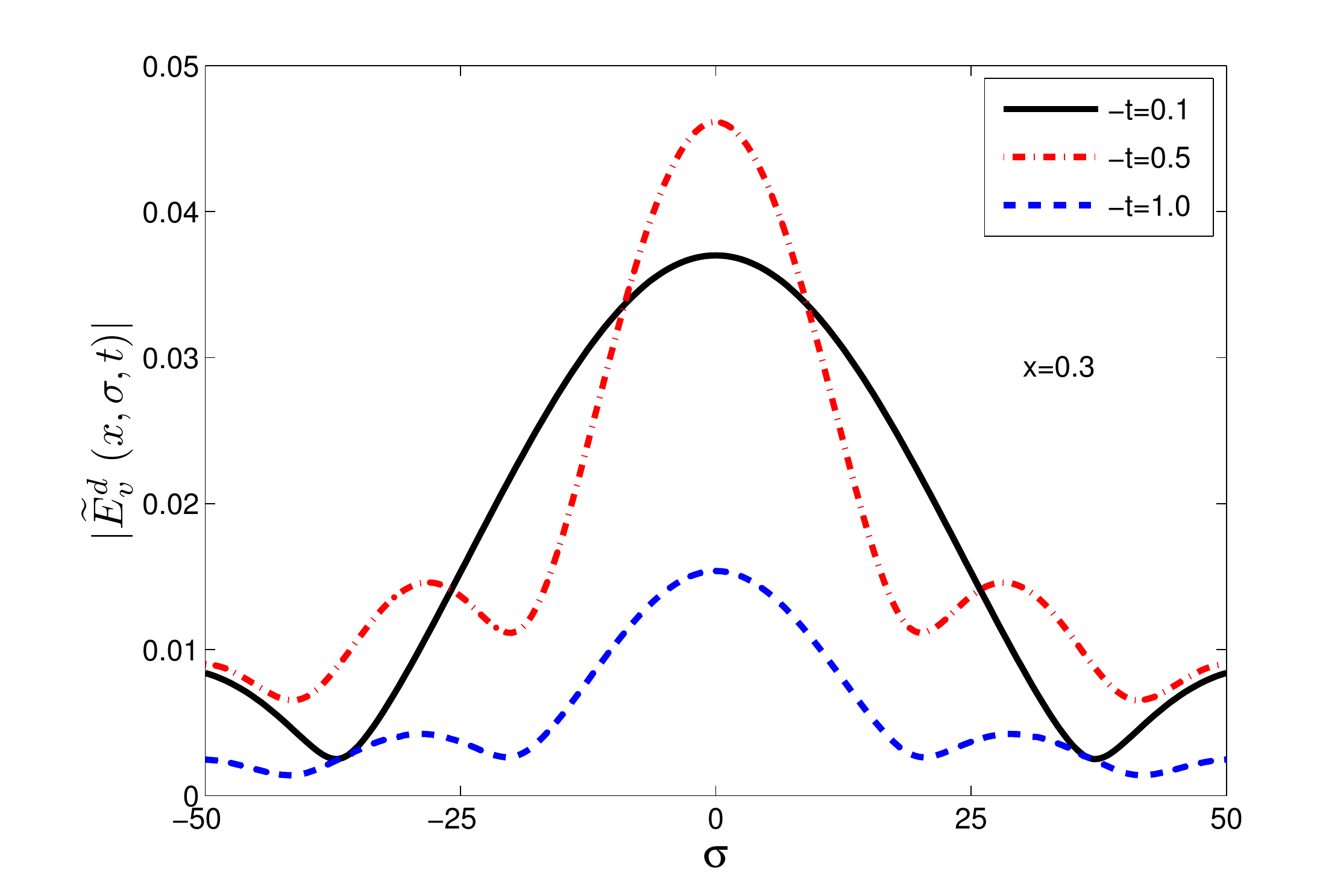}
\end{minipage}
\caption{\label{glx6}(Color online) Plots of the helicity dependent GPDs in longitudinal impact space vs $\sigma$ and different values of $-t$ in $\rm{GeV}^2$, for fixed value of $x=0.3$. Left panel is for $u$ quark and the right panel is for $d$ quark.}
\end{figure*}
The Fourier transform of GPDs with respect to the skewness variable $\zeta$ provides a unique way to visualize the structure of the hadron in the boost-invariant longitudinal coordinate space.
The boost invariant longitudinal impact parameter is defined as $\sigma=\frac{1}{2}b^-P^+$ which was first introduced in \cite{BDHAV}. It has been shown that the DVCS amplitude in a QED model of a dressed electron exhibits an interesting diffraction pattern in the longitudinal impact parameter space in analogous to diffractive scattering of a wave in optics~\cite{BDHAV}.  
The finite size of the $\zeta$ is responsible for producing the diffraction pattern and this can be interpreted as a slit of finite width in equivalent with optics. We should mentioned here that the Fourier transform with a finite range of $\zeta$ of any arbitrary function does not provide the diffraction pattern~\cite{CMM2}. This pattern depends on the nature of the function. The helicity dependent GPDs for photon evaluated in a phenomenological model \cite{Mukherjee:2011an} show similar diffraction pattern in the longitudinal impact parameter space. A phenomenological model for proton GPDs also exhibits the similar diffraction pattern \cite{CMM2} whereas the GPDs calculated for a simple relativistic spin half system of an electron dressed with a photon display a same pattern in the longitudinal position space \cite{CMM1,Kumar1}. The  similar phenomenon are also observed for the unpolarized GPDs as well as chiral-odd GPDs in this light front quark-diquark model~\cite{Mondal:2015uha,Chakrabarti:2015ama}.
\begin{figure*}[htbp]
\begin{minipage}[c]{0.98\textwidth}
\small{(a)}
\includegraphics[width=7.5cm,height=5.15cm,clip]{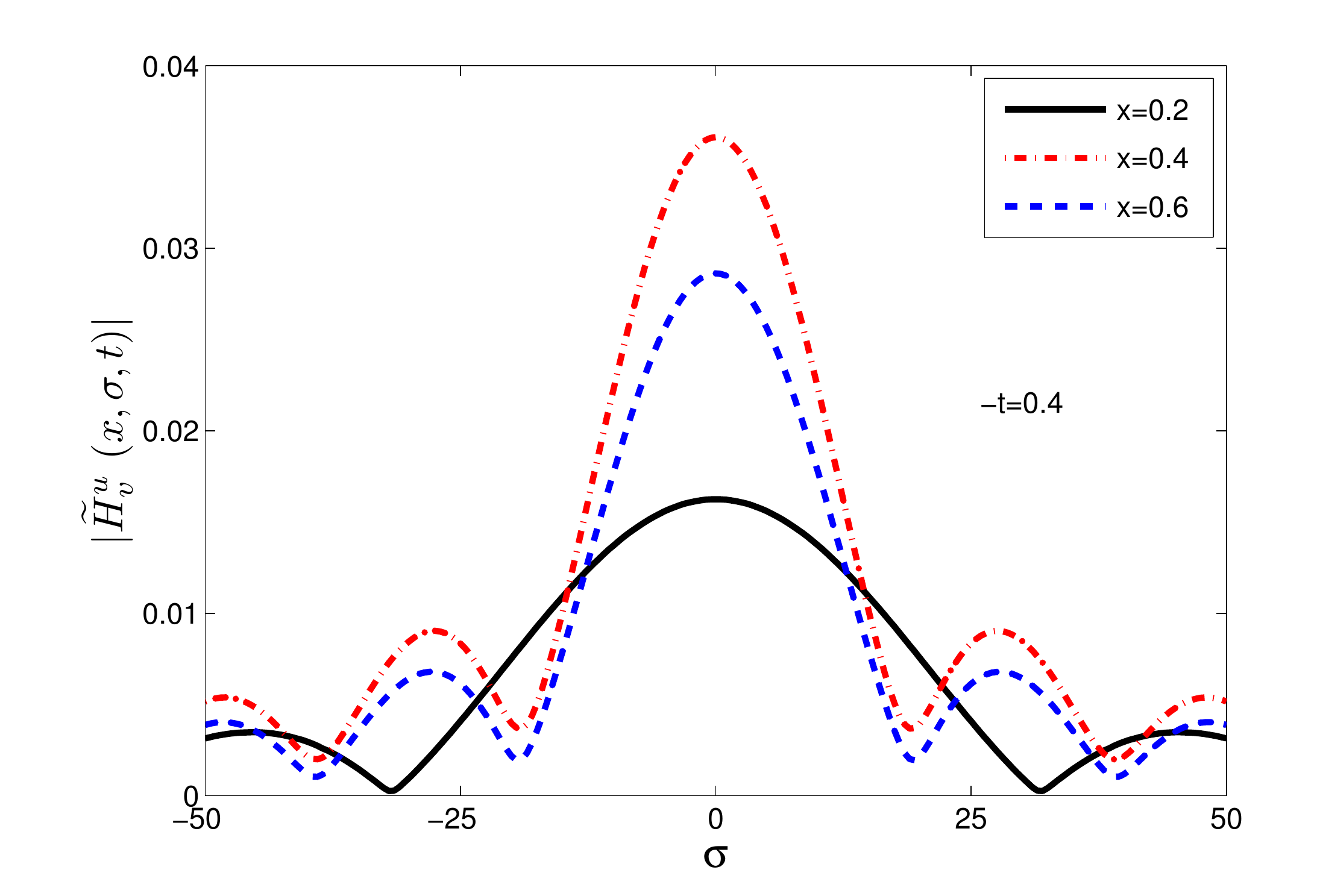}
\hspace{0.1cm}%
\small{(b)}\includegraphics[width=7.5cm,height=5.15cm,clip]{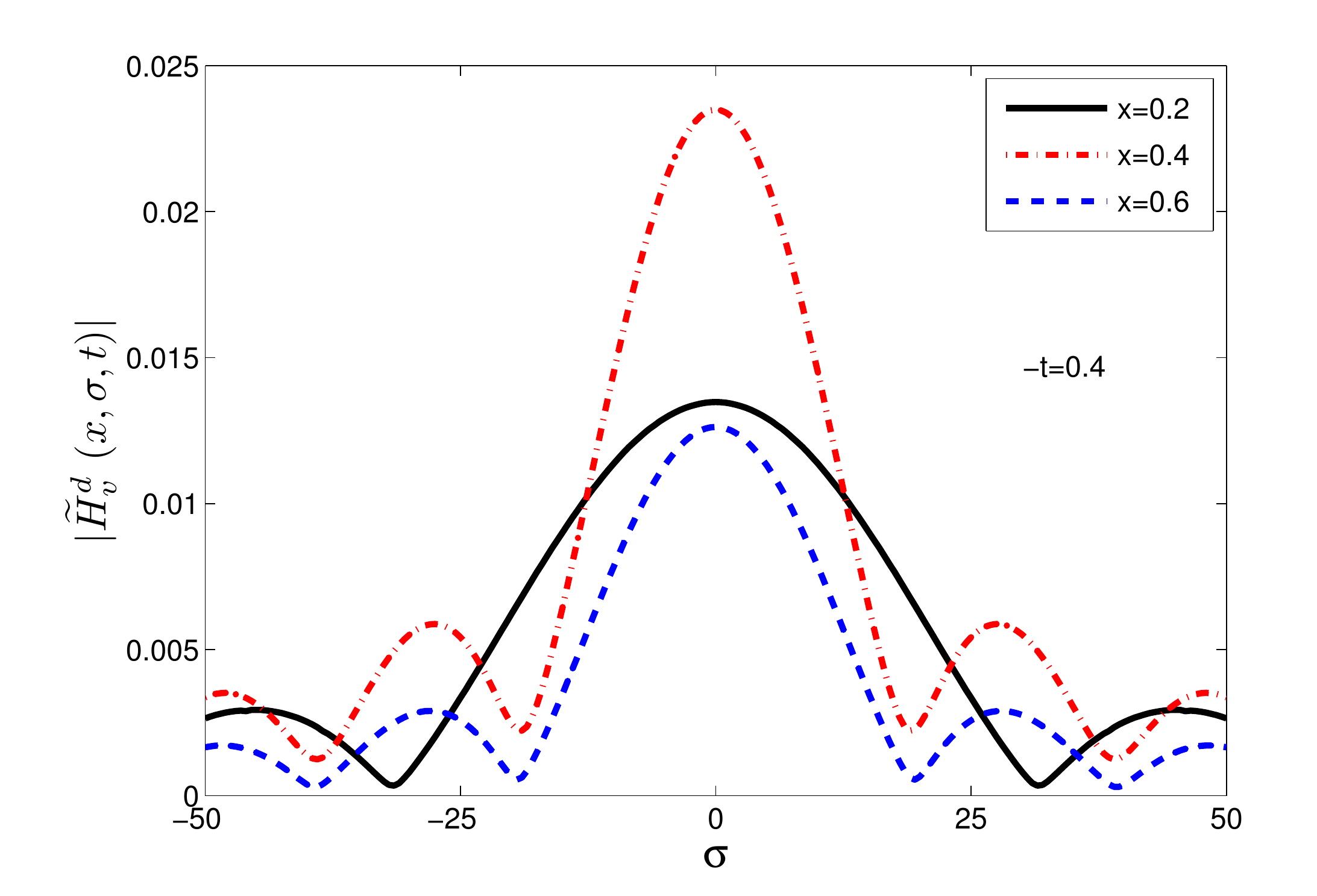}
\end{minipage}
\begin{minipage}[c]{0.98\textwidth}
\small{(c)}\includegraphics[width=7.5cm,height=5.15cm,clip]{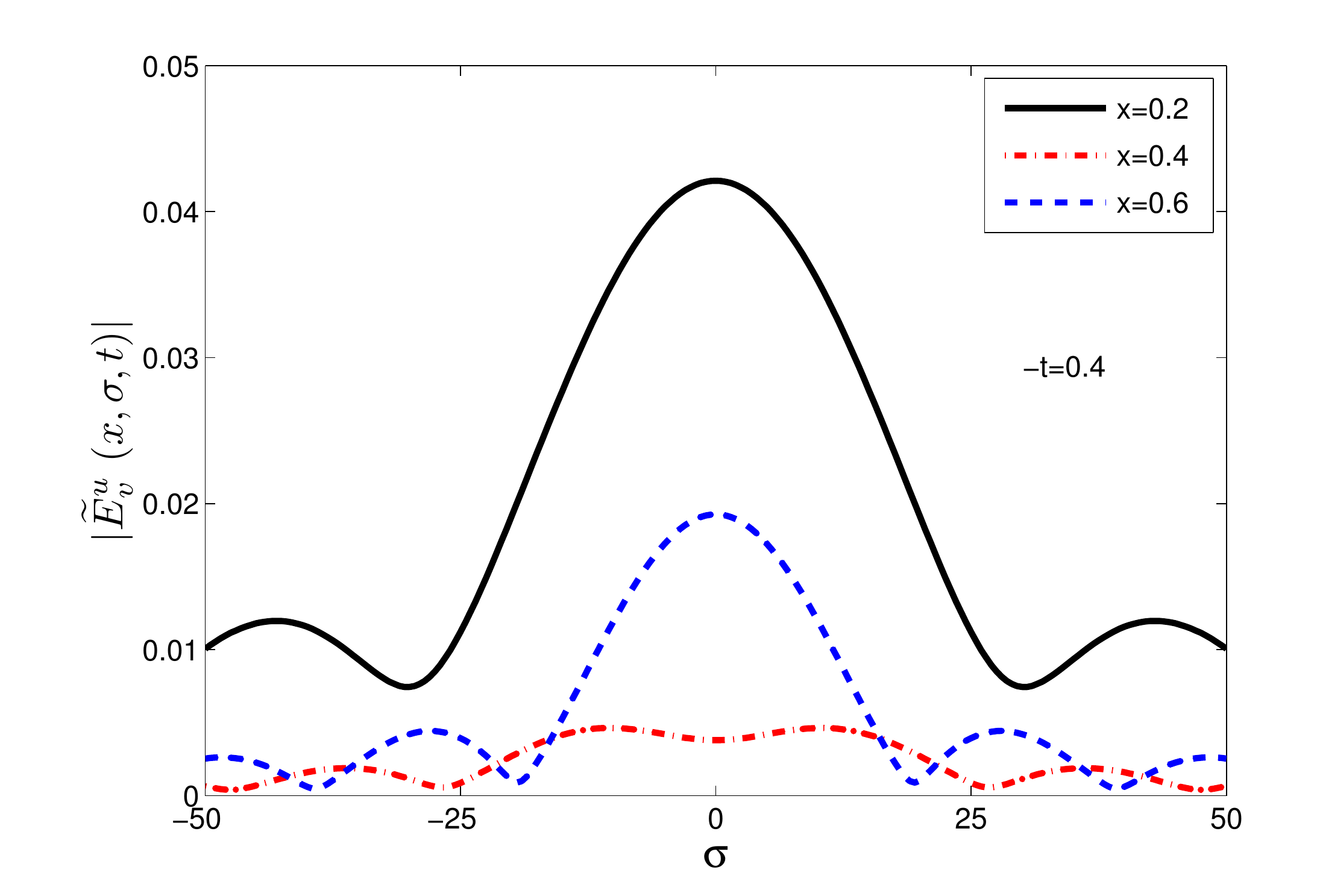}
\hspace{0.1cm}%
\small{(d)}\includegraphics[width=7.5cm,height=5.15cm,clip]{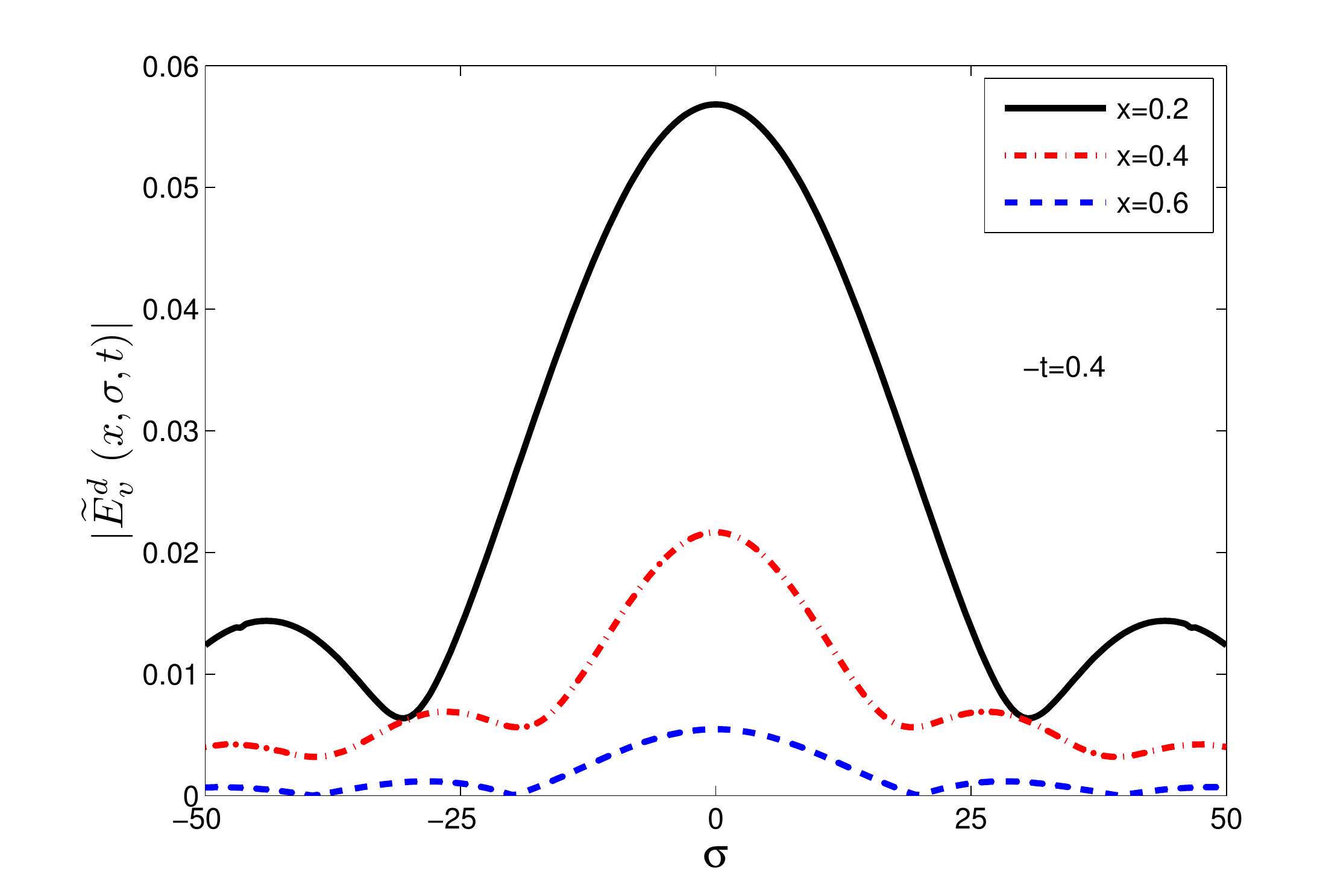}
\end{minipage}
\caption{\label{glt5}(Color online) Plots of the chiral-odd GPDs in longitudinal impact space vs $\sigma$ and different values of $x$, for fixed value of $-t=0.4$ $GeV^2$. Left panel is for $u$ quark and the right panel is for $d$ quark. For $-t=0.4$ $GeV^2$, $\zeta_{max}\approx0.307$.}
\end{figure*}
In longitudinal position space, the GPDs are defined as
\be
\widetilde{H}(x,\sigma,t)&=&{1\over 2 \pi} \int_0^{\zeta_f} d \zeta e^{i\zeta P^+b^-/2}\widetilde{H}(x,\zeta,t),\nonumber\\
&=&{1\over 2 \pi} \int_0^{\zeta_f} d \zeta e^{i\zeta \sigma}\widetilde{H}(x,\zeta,t),\\
\widetilde{E}(x,\sigma,t)&=&{1\over 2 \pi} \int_0^{\zeta_f} d \zeta e^{i\zeta P^+b^-/2}\widetilde{E}(x,\zeta,t),\nonumber\\
&=&{1\over 2 \pi} \int_0^{\zeta_f} d \zeta e^{i\zeta \sigma}\widetilde{E}(x,\zeta,t).
\ee
Since the region of our discussion is $\zeta<x<1$, the upper limit of $\zeta$ integration, $\zeta_f$ is given by $\zeta_{max}$ if $x$ is larger than $\zeta_{max}$, otherwise by $x$ if $x$ is smaller than $\zeta_{max}$ where the maximum value of $\zeta$ for a fixed $-t$ is given by
\be
\zeta_{max}=\sqrt{\frac{(-t)}{(-t+4M_n^2)}}.
\ee
The Fourier spectrum of the helicity dependent GPDs for $u$ and $d$ quarks in longitudinal position space as a function of $\sigma$ for different values of $-t$ and fixed $x=0.3$ are shown in Fig.\ref{glx6}. $\widetilde{H}$ for both $u$ and $d$ quarks  displays a diffraction pattern in the $\sigma$ space as observed for the DVCS amplitude~\cite{BDHAV}. We also observe that $\widetilde{E}(x,\sigma,t)$ for $d$ quark exhibits a same pattern but for all values of $-t$, it does not show the prominent pattern for $u$ quark. This is due to the fact that the distinctly different nature of $\widetilde{E}^u(x,\zeta,t)$ with $\zeta$ compared to the other GPDs which again implies that the diffraction pattern is not solely due to the finite size of $\zeta$ integration, the functional form of the GPDs are also important for this phenomenon. The first minima appears at the same values of $\sigma$ for all the diffraction patterns. In Fig.\ref{glt5}, we also show the GPDs in $\sigma$ space for different values of $x$ and fixed $-t=0.4$ $GeV^2$. Here $\zeta_{f}$ plays the role of the slit width in equivalent to the single slit optical diffraction pattern.  Since the positions of the minima are inversely proportional to the slit width, as the slit width $\zeta_{f}$ increases, the minima sifts towards the center of the diffraction pattern. 
\begin{figure*}[htbp]
\begin{minipage}[c]{0.98\textwidth}
\small{(a)}
\includegraphics[width=7.5cm,height=5.15cm,clip]{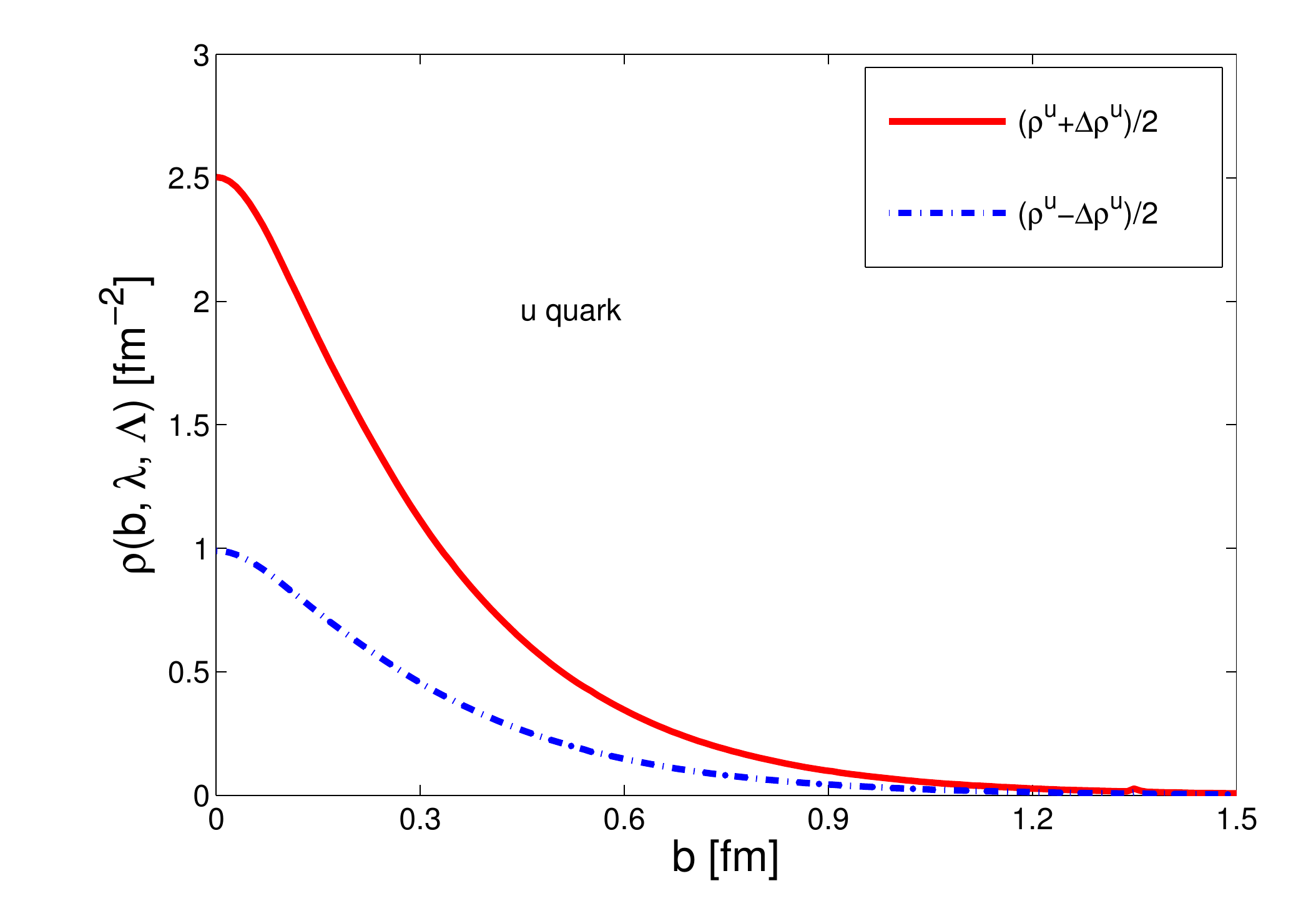}
\hspace{0.1cm}%
\small{(b)}\includegraphics[width=7.5cm,height=5.15cm,clip]{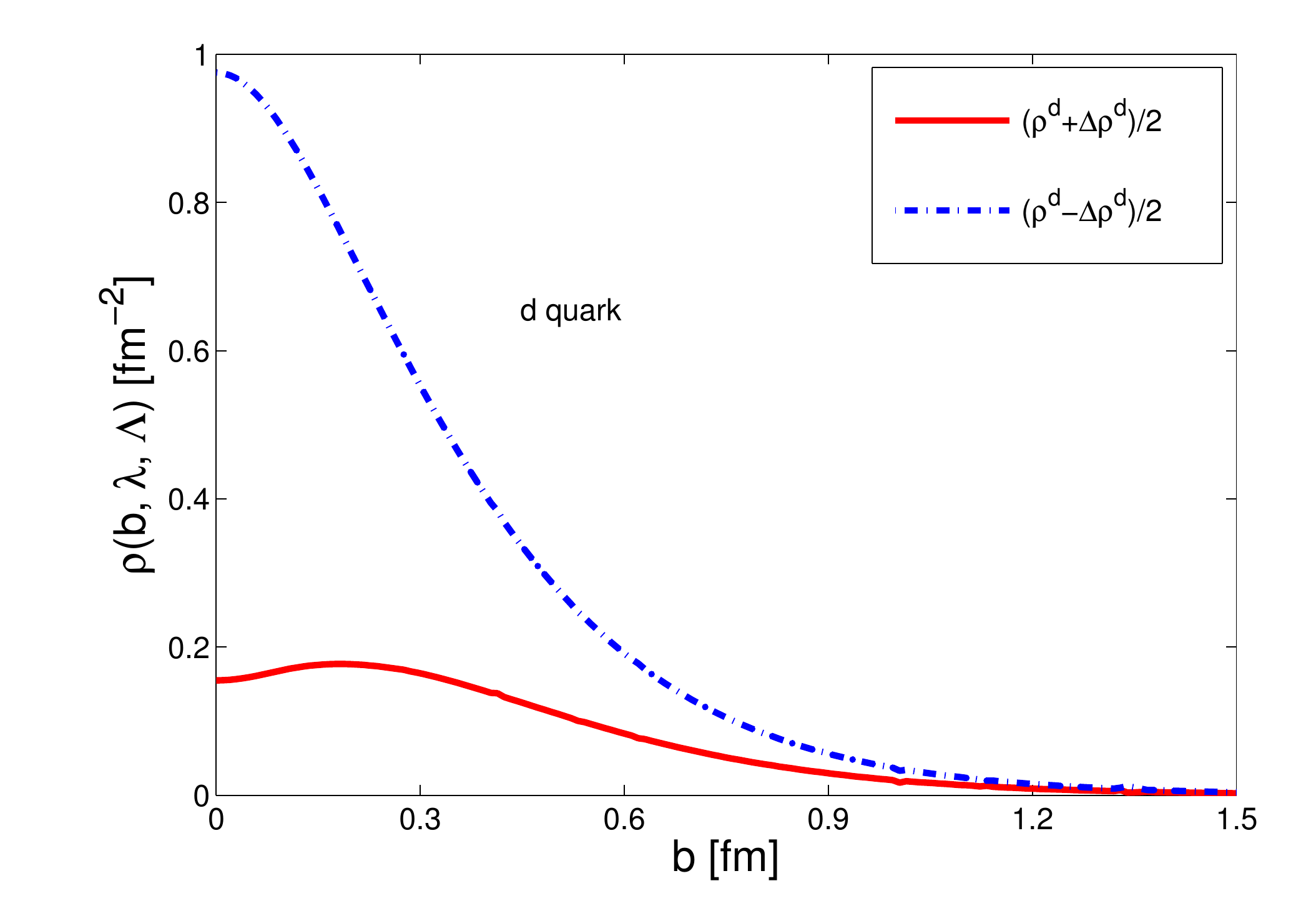}
\end{minipage}
\begin{minipage}[c]{0.98\textwidth}
\small{(c)}\includegraphics[width=7.5cm,height=5.15cm,clip]{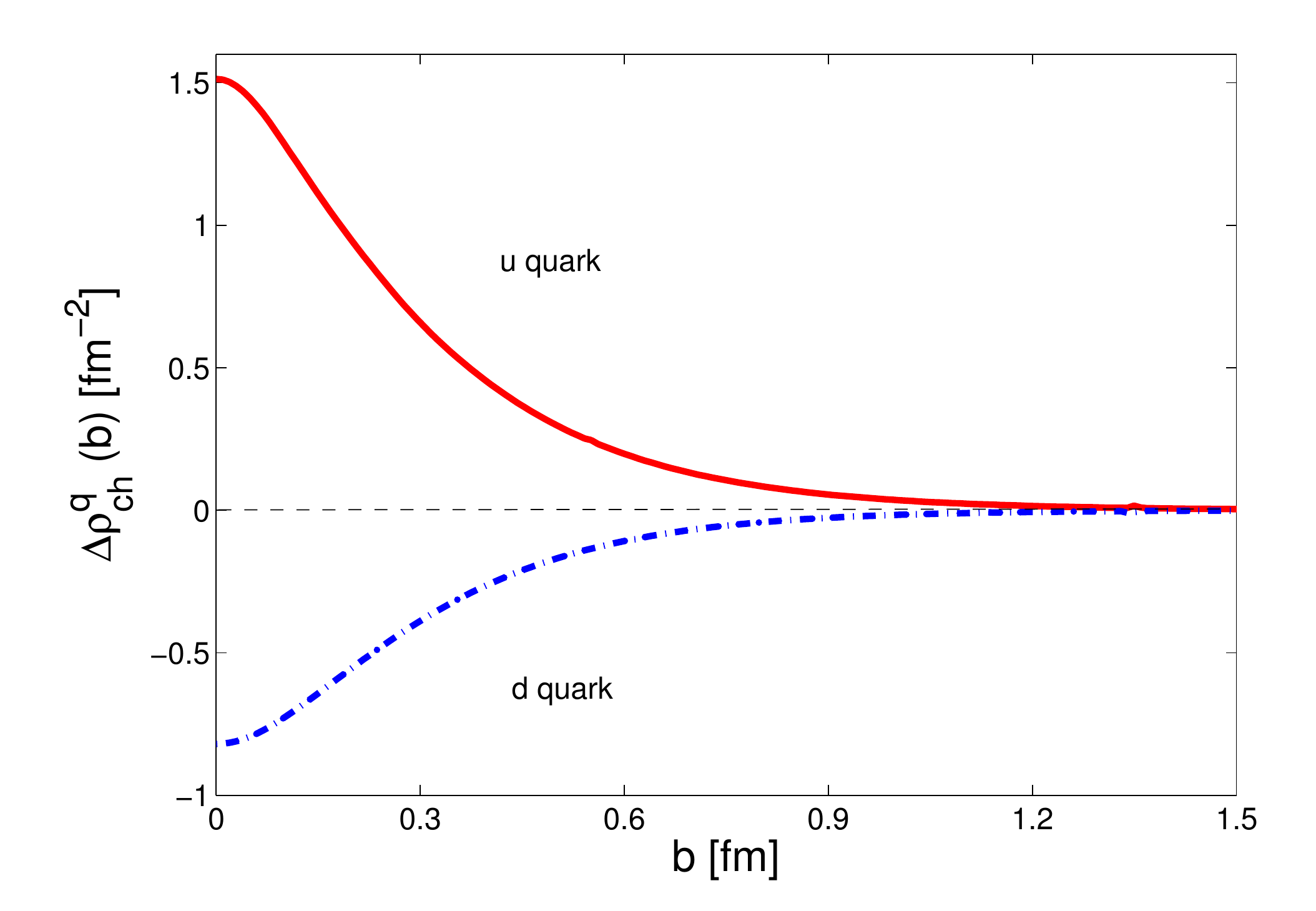}
\hspace{0.1cm}%
\small{(d)}\includegraphics[width=7.5cm,height=5.15cm,clip]{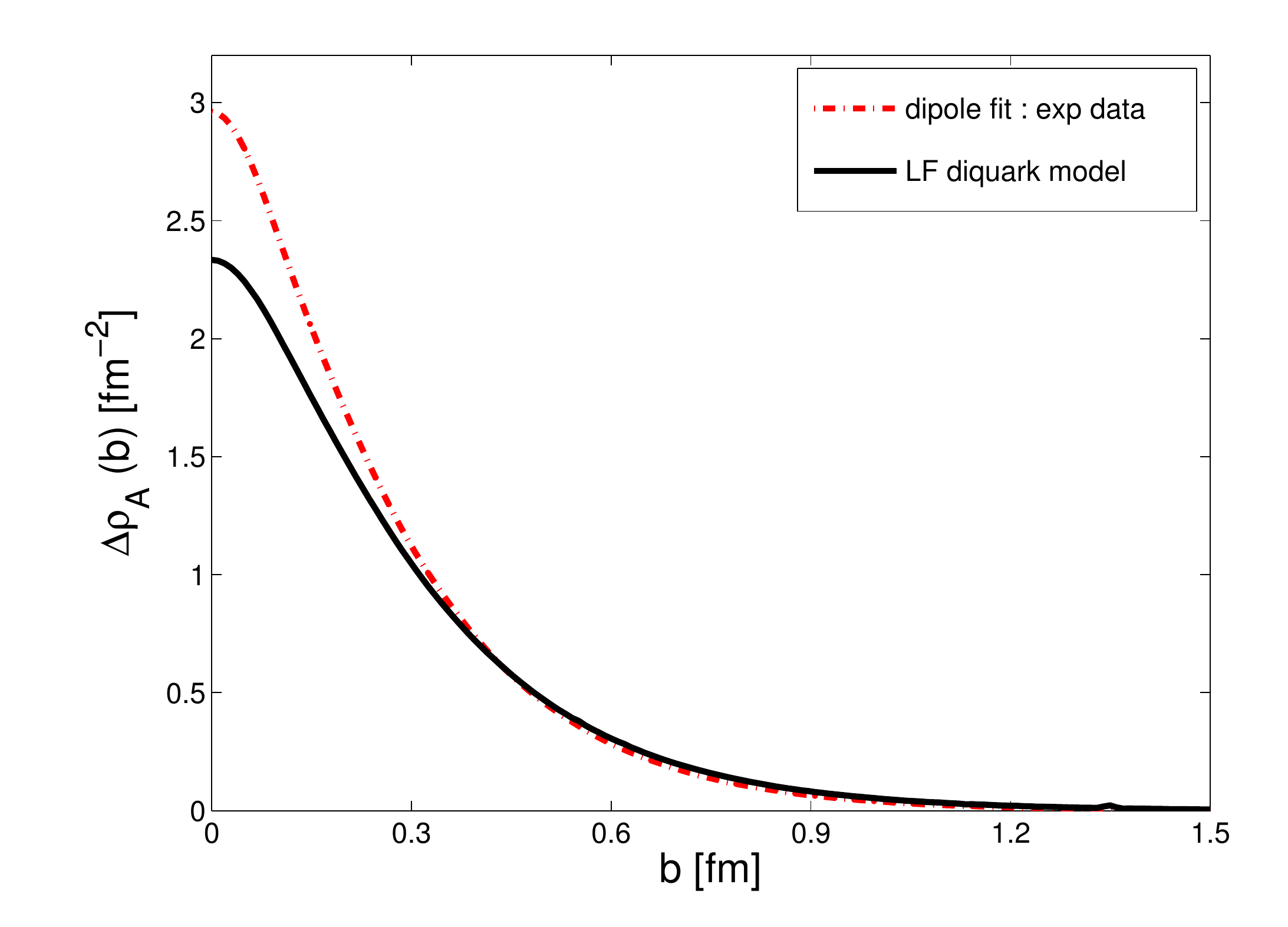}
\end{minipage}
\caption{\label{densities}(Color online) Plots of the transverse distribution of $u$ and $d$ quarks in a longitudinally polarized proton as a function of the impact parameter $b$. Total contribution for (a) $u$ quark, and (b) $d$ quark when quarks are polarized in the longitudinal
direction, either parallel (solid red lines) or anti-parallel (dashed blue lines) with respect to the proton helicity. (c) The axial contributions $\Delta u$ and $\Delta d$ for $u$ and $d$ quarks. (d) The axial distribution $\rho_A(b)=\Delta u(b)-\Delta d(b)$ (solid black line) in comparison with the distribution from the dipole fit of experimental data for axial form factor (red dashed dot). }
\end{figure*}
\subsection{Quark transverse distributions}\label{quark_distribution}
 One can access the probability $\rho^q(b,\lambda,\Lambda)$ to find a quark with transverse position $b$ and light-cone helicity $\lambda$ ($=\pm 1$) in the nucleon with longitudinal polarization $\Lambda$ ($=\pm 1$) via Fourier transform of the combination of the Dirac and axial form factors of quark as~\cite{Pasquini1,Pasquini2,Pasquini3},
\be
\rho^q(b,\lambda,\Lambda) &=& \frac{1}{2}\int {\rm d}^2\mathbf{\Delta}_\perp
\left[ F^q_1(Q^2) +\lambda\Lambda G^q_A(Q^2) \right]\,e^{i\mathbf{\Delta}_\perp\cdot\mathbf{b}_{\perp}} \nonumber \\
&=& \frac{1}{4\pi}\int {\rm d}Q \,Q J_0(Qb) \left[ F^q_1(Q^2) + \lambda\Lambda G^q_A(Q^2) \right] \nonumber \\
&\equiv& \frac{1}{2}\left[\rho^q(b) +  \lambda\Lambda \Delta q(b)\right],
\ee
where $\rho^q(b)$ and $\Delta q(b)$ are the Fourier transform of $F^q_1(Q^2)$ and $G^q_A(Q^2)$ respectively and $J_0$ is a cylindrical Bessel function. $\rho^q(b)$ corresponds to $d(b)$, the charge density for $d$ quark and $2u(b)$,  twice charge density for $u$ quark \cite{Mondal:2015uha,Pasquini3}. The normalization of $\rho^q(b)$ and $\Delta q(b)$ are: $\int {\rm d}^2b~\rho^q(b)=n_q$, where $n_u=2$, $n_d=1$ in proton and $\int {\rm d}^2b ~\Delta q(b)=\Delta q$, where $\Delta q$ is the axial charge of quark $q$.
We show the resulting probability for $u$ and $d$ quarks considering a positive proton helicity ($\Lambda=1$) in Fig.\ref{densities}(a) and Fig.\ref{densities}(b), respectively. The axial contributions $\Delta u (b)$ and $\Delta d (b)$ for $u$ and $d$ quarks having opposite sign are shown in Fig.\ref{densities}(c) whereas the transverse distribution $\rho^q(b)$, which is positive for both $u$ and $d$, in this light-front quark diquark model can be found in \cite{Mondal:2015uha}. The difference between $\Delta u (b)$ and $\Delta d (b)$ is compared with the distribution obtained from the dipole fit of axial form factor in Fig.\ref{densities}(d). One can notice that though there is a mismatch at $b=0$, at larger $b$, light-front quark-diquark model agree well the result obtained from dipole fit. Since $\Delta u (b)$ is positive but $\Delta d (b)$ is negative, the probability to find a $u$ quark with positive helicity is maximal when it is aligned with the proton helicity while the opposite
occurs for $d$ quarks.

\section{Summary}\label{summary}
In the present work, we have studied the helicity dependent GPDs for $u$ and $d$ quark in proton for nonzero skewness in the light front quark-diquark model predicted by the soft-wall AdS/QCD. We have obtained the GPDs in terms of the overlaps of the light-front wavefunctions considering the DGLAP region i.e., for ($x>\zeta$). We have observed that for fixed $\zeta$ The peaks of the distributions move to higher values of $x$ with increasing of $-t$ again the height of the peaks increases and also shift to higher values of $x$ as $\zeta$ increases for fixed  $-t$. We also observed markedly different behavior for $\widetilde{E}$ for $u$ quark from the other GPDs in this model when we plot the GPDs against $\zeta$ for different $-t$ and fixed $x$. It shows that with increasing $\zeta$, $\widetilde{E}^u$ started to increase smoothly from different values at $\zeta=0$ for different values of $-t$ but the magnitude at $\zeta_{max}$ decreases with increasing $-t$ whereas for the other GPDs, the magnitude at $\zeta_{max}$ increases with increasing $-t$. The axial form factor has been evaluated in this quark-diquark model and compared with the dipole fit of experimental data as well as lattice data. It shows that our result is more or less in agreement with the experimental data and better compared to lattice. 

We have also presented all the helicity dependent GPDs in the transverse impact parameter($b$) as well as longitudinal position($\sigma$) spaces by taking the Fourier transform of the GPDs with respect to momentum transfer in the transverse direction (${\bf \Delta}_{\perp}$), and skewness ($\zeta$), respectively. For zero skewness, the impact parameter $b$ gives a measure of  the transverse distance between the struck parton and the center of momentum of the hadron.
In this model, the GPD $\widetilde{H}$ shows quite different behavior in the transverse impact parameter space for $u$ and $d$ quarks when plotted in $x$ and $b$ but for  $\widetilde{E}$, the behaviors for both $u$ and $d$ quark are almost same. Again, the nature of $\widetilde{H}$ are more or less same when plotted against $\zeta$ and $b$ but $\widetilde{E}$ shows a different behavior for $u$ and $d$ quark. With increasing $\zeta$ or decreasing $x$, the width of the all distributions increase. It has been found that the GPDs  in $\sigma$ space show diffraction patterns analogous to diffractive scattering of a wave in optics. 
A similar diffraction pattern also has been  observed  in several other models. The qualitative nature of the diffraction patterns are same for both $u$ and $d$ quarks.
The general features of this phenomenon are mainly depending on the finiteness of $\zeta$ integration but the dependence of GPDs on $x$, $\zeta$ and $t$ is also crucial. Like other GPDs, $\widetilde{E}$ for $u$ quark does not show the diffraction pattern for all values of $-t$. This is due to a different behavior of $\widetilde{E}^u$ with $\zeta$ from the other GPDs which also indicates that the diffraction pattern is  not solely  due to finiteness of $\zeta$ integration and the functional behaviors of the GPDs are  important to have the phenomenon.
In this model, we have also studied the transverse distributions of quark with light-cone helicities $\lambda(=\pm 1)$ in the nucleon with longitudinal polarization $\Lambda(=+1)$. We observed that when the helicity of $u$ quark is aligned with the proton helicity, the probability to find it is maximal but the situation is opposite for $d$ quark. 




\end{document}